# Multi-Scenario Empirical Assessment of Agile Governance Theory: A Technical Report

Alexandre J. H. de O. Luna, Ph.D.

Department of Management Sciences (DCA), Center for Applied Social Sciences (CCSA). Federal University of Pernambuco (UFPE). dos Funcionários Avenue. ZIP Code: 50.740-580. Recife, Pernambuco, Brazil. *alexandre.luna@ufpe.br*

Informatics Center (CIn). Federal University of Pernambuco (UFPE). Jornalista Anibal Fernandes Avenue. ZIP Code: 50.732-970. Recife, Pernambuco, Brazil. ajhol@cin.ufpe.br

Marcelo L. M. Marinho, Ph.D.

Department of Computer Science (DC). Federal Rural University of Pernambuco (UFRPE). Dom Manoel de Medeiros Street. ZIP Code: 52171-900. Recife, Pernambuco, Brazil. *marcelo.marinho@ufrpe.br*

──────────────── ◆ ────────────────

**ABSTRACT - Context**: Agile Governance Theory (AGT) has emerged as a potential model for organizational chains of responsibility across business units and teams. **Objective**: This study aims to assess how AGT is reflected in practice. **Method**: AGT was operationalized down into 16 testable hypotheses. All hypotheses were tested by arranging eight theoretical scenarios with 118 practitioners from 86 organizations and 19 countries who completed an in-depth explanatory scenario-based survey. The feedback results were analyzed using Structural Equation Modeling (SEM) and Confirmatory Factor Analysis (CFA). **Results**: The analyses supported key theory components and hypotheses, such as mediation between agile capabilities and business operations, through governance capabilities. **Conclusion**: This study supports the theory and suggests that AGT can assist teams in gaining a better understanding of their organization's governance in an agile context. A better understanding can help remove delays and misunderstandings that can come about with unclear decision-making channels, which can jeopardize the fulfillment of the overall strategy.

──────────────── ◆ ────────────────

## 1. INTRODUCTION

While *governance* denotes how an organization is governed, including decision-making structures and controls (Moe et al., 2021), *agility* is focused on reacting and adapting rapidly to changes, and *lean*[1] is focused on combating wastage. Both agility and lean are intrinsically related to delivering value, although these approaches may sometimes conflict. Wang et al. (2012) suggest that a rational balance between the two can result in a unified "agile" approach that can achieve better results than if applied separately. We adopt this merged approach for agile. *Agility* and *lean thinking* stand out as relevant approaches to some domains beyond agile software development (Sakhrawi et al., 2022), such as emergent technologies management to provide business flexibility and security (Pal et al., 2021). As a survival instrument for firms in the market (Škare and Soriano, 2021). In public management, by seeking to establish mechanisms for a responsive government (Balakrishnan et al., 2022) and develop dynamic capabilities in public organizations (Panagiotopoulos et al., 2022). By dealing with customers' expectations in the automotive sector (Giacosa et al., 2022) and new products development (Tseng et al., 2022). In enterprise social media, by aiming to stimulate employee agility (Pitafi et al., 2020). In healthcare, by pointing out lean management principles to promote socially responsible innovation in the U.S. healthcare system (Batayeh et al., 2018). Also impacting on how startups and incumbents innovate their business models to reduce uncertainty, engage stakeholders, and promote collective learning (Bocken and Snihur, 2020). Moreover, in fusing financing and technology management by orchestrating all techno-financing systems and strategies for rapid enterprise growth (Tou et al., 2020).

The *Agile Governance* (AG) idea emerged initially as a deterministic approach to guide software development (Qumer, 2007), evolving into an adaptive and reflexive approach focused on organizational performance, competitiveness, and sustainability, whose application occurs in different areas (Founoun et al., 2022). AG is conceptualized as "*the capability of an organization to sense, adapt and respond to changes in its environment, in a coordinated and sustainable way, faster than the rate of these changes*" (Luna et al., 2016).

Although governance drives organizational performance, its intrinsic controls might limit an organization's ability to adapt to change quickly. Considering these issues, the Agile Governance Theory (AGT) emerged years ago, aiming to analyze and describe phenomena related to how teams can develop the intrinsic dynamic capabilities[2] to sense and respond to organizational or requirement changes (Luna et al., 2020). AGT seeks to prepare teams to respond and even anticipate those changes in a coordinated and sustainable manner. AGT also provides mechanisms for describing and analyzing the factors and agents that

───────

[1] Lean is a mindset that involves never-ending efforts to eliminate or reduce *"muda"* (the Japanese word for waste) and deliver value. "Lean thinking" could be synthesized as an attitude of "doing more with less."

[2] The term *"capability"* relates to a feature, *faculty* or process that can be developed or improved (Vincent, 2008).

influence agile governance practice in organizations, which can often remain hidden or difficult to notice. However, the assessment of the theory in practice remains unexplored.

This report assesses whether the Agile Governance Theory (AGT) reflects practitioners' experience. We imply that once we have tested the AGT key hypotheses and core assumptions, we can achieve better reliability in using the theory to identify strengths, weaknesses, opportunities, and threats in how organizations deal with fast-changing competitive environments. In this report, we test the extent to which the whether the behaviors predicted by the AGT are reflected in the experience of practitioners by applying an assessment mechanism developed by (Luna et al., 2020) in practice and ask:

> *RQ: Are the key hypotheses and core assumptions of the Agile Governance Theory supported by practitioners' experience?*

This article is structured as follows. **Section 2** provides an overview of agile governance theory. In **Section 3,** we describe the methods employed. In **Section 4,** we present the results of this study. **Section 5** addresses the research question and discusses the strength of evidence, indications for research and practice, and study limitations. **Section 6** concludes the study and considers opportunities for further research on agile governance.

## 2. BACKGROUND

Agile Governance (AG) has evolved, being a concept that emerged in Software Engineering (Qumer, 2007), and explored in the software product lines context (Cheng et al., 2009; Dubinsky et al., 2008). Then it started to be applied within the scope of Information Technology Governance (ITG) (Luna et al., 2010) as an approach to direct and manage technological products, services, and companies to provide business agility.

Recent works denote the increase in the range of applications of AG. Under the agile services paradigm, Maurio et al. (2021) seek to use intelligent agents in a modeling and simulation framework to test the resiliency of autonomous unmanned aerial systems and to measure the efficiency and safety of their operations in a simulated multi-UAS air-traffic control context. In the Banking industry, AG influencing factors emerge from pondering enablers, hindering variables, and barriers to adopting and managing emerging technologies that can boost business operations and better deliver organizational values for the core business (Saheb and Mamaghani, 2021). In the context of international high-tech small and medium-sized enterprises (SMEs), by seeking to develop dynamic capabilities to assist high-tech SMEs in becoming agile in their cross-border activities (Jafari-Sadeghi et al., 2022). In the context of Global Business, in the practice of sustainability in supply chains for multinational companies in emerging markets (Soundararajan et al., 2021). In Education, by proposing an agile approach that emphasizes self-direction, collaboration, and lightweight procedures, for fostering innovation in university teaching and learning (Wirsing and Frey, 2021). As input for the elaboration of instruments for Digital Governance, aiming to improve detection, decision, and response capabilities in turbulent business environments (Vaia et al., 2022). In the Smart Cities context, by favoring the applications of new technologies for urban planning and facilitating citizens to actively interact with decision-makers (Founoun et al., 2022; Hahn and te Brömmelstroet, 2021). Alternatively, even by guiding to overcome challenges and deal with dynamic environmental conditions during the Covid-19 pandemic (Halim et al., 2021; Janssen and Voort, 2020).

Although Agile Governance is being applied more and more in distinct areas, the related phenomena have not yet been explored in depth until then. Aiming to shed light on the nature of phenomena related to Agile Governance, the Agile Governance Theory (AGT) emerged (Luna, 2015) as an instrument to analyze and describe those phenomena.

To demonstrate AGT applicability let us analyze the concept of "Ambidextrous Governance", sometimes confused with Agile Governance. According to O'Reilly and Tushman (2004), an "ambidextrous organization" is presented as a *modus-operandi* in which organizations can develop disruptive innovations and generate new competitive advantages while maintaining their ability to operate their traditional businesses. It should happen through integrating the roles of managers, entrepreneurs, and leaders and is focused on the separation of processes, structure, and culture between the emerging structures involved in an innovation context and the pre-existing traditional organizational structures, managed through a tightly integrated senior team to achieve operational resilience (Iborra et al., 2020). In this context, "Ambidextrous Governance" is described as a dual governance model in which firms alternate between traditional and agile ITG mechanisms (Vejseli et al., 2022). On the other hand, Agile Governance aims to influence the whole steering system of an organization, responsible for the perception, response, and coordination of every component of the corporate body. Under the lens of AGT (Luna et al., 2020), in terms of agility, considering the previous characterization, Ambidextrous Governance would be classified as a 'specific agile approach' because it limits its influence to a localized outcome or a stage in the organization's value chain (Porter, 1985). After all, in this ambidextrous approach, part of the organization would not be within the agile scope of influence.

AGT advocates that the development of agility and governance capabilities by the organization should occur iteratively and incrementally. So, processes, structures, and cultures can co-exist while the organization is evolving and developing such capabilities. However, this eventual operational redundancy must occur in a transitory way. In the long run, the existence of different cultures within the same organization can cause more harm than good. For instance, considering ITG mechanisms, AGT advocates that the mechanisms be agile and resilient. At the same time, they must be compliant because compliance is not a matter of choice but conformity and necessity.

AGT also proposes an approach based on an analytical and reflexive balance on the combined use of agile and lean capabilities: prompting resilience, adaptability, and speed of response (effectiveness), in a coordinated way, seeking to minimize waste (efficiency) during organizational transformations resulting from those adaptations and response to change. AGT advocates that, sometimes, it is necessary to consider some situations in which the team needs to be predominantly agile, seeking to react and adapt to change, even if it generates some future rework (waste) so as not to lose market timing. While eventually, other contexts may require the team to be predominantly lean, seeking to minimize rework, react more slowly, and adapt to change progressively to avoid waste (rework).

AGT characterizes 'agile governance' as a socio-technical phenomenon where people must be agents of change in organizational contexts, where technology is often a crucial transformation factor. The AG's socio-technical nature also derives from the need to understand and handle the intersections between technical and social aspects, enabling decision-makers to thoughtfully and intentionally deal with the social forces that shape technological decisions and the choices that are open to society concerning technology use.

AGT recognizes the value of (1) behavior and practices over (A) process and procedures, (2) achieving sustainability and competitiveness over (B) being audited and compliant. Besides, AGT values (3) transparency and people's engagement over (C) their monitoring and controlling, as well as (4) organizational abilities for sensing, adapting, and responding over (D) than just following a pre-arranged plan. Although (A), (B), (C), and (D) are non-negotiable aspects, as they are essential instruments for good governance, AGT considers that (1), (2), (3), and (4) are crucial transformation factors for sustainable business agility.

Considering AGT as a context-sensitive theory, it can be instantiated in different *units of analysis*[3], called *organizational context*[4], in which the choice of the unit of analysis depends on the organizational environment we intend to analyze and describe. An *organizational context* is described by the collective behavior of the agents contained within it. For instance, AGT might be instantiated in the following organizational contexts: teams, projects, business units, enterprises, or even in a multi-organizational setting.

AGT was conceptualized (Luna et al., 2015) as a system characterized by the behavior of six theoretical constructs that seek to describe and explain agile governance phenomena through their relationships and interactions, named: *Environmental factor effects[E], Effects of moderator factors [M], Agile capabilities [A], Governance capabilities [G], Business operations [B],* and *Value delivery [R]*. They are briefly characterized, in keeping with, as follows:

- **Effects of environmental factors [E]** depict the effects sensed in the organizational context because of the influence of the external environment in which the organizational context resides. For instance, regulatory effects derived from legal, economic, or political issues; technological obsolescence or innovation; market competitiveness or turbulence; economic growth or decline, among others.

- **Effects of moderator factors [M]** typify the effects sensed in the organizational context because of the influence of inhibitory or restrictive factors that form part of this context, e.g., organizational culture; absence of leadership or oppressive leadership; facilities or complications resulting from the organizational architecture or business model; people qualifications, and motivation.

- **Agile capabilities [A]** represent the ability to acquire, develop, apply, and evolve competencies related to rapidly and adaptively addressing changing environments, considering the principles, values, and practices of agile and lean philosophy in the organizational context, such as flexibility, adaptability, and leanness.

- **Governance capabilities [G]** identify the ability to acquire, develop, apply, and evolve dynamic competencies related to how an organizational context is conducted, administered, or controlled, including the relationships between the different parties involved and the aims for which it is governed (e.g., processes, policies, laws, customs, and institutions).

- **Business operations [B]** characterize a set of organized activities that are part of the day-to-day business functions, conducted to generate value delivery, including (but not limited to): processes, functions, services, products, projects, practices, and behaviors.

- **Value delivery [R]** portrays the ability to generate results for the business by delivering value, including all forms of value that determine the health and well-being of the organization in the long run. For instance, what can be perceived through the "quality" of products and services, which in turn, can be broken down into dimensions: "utility" (fit for purpose) and "warranty" (fit for use); as well as the "time-to-market" to deliver those products and services.

Besides the constructs, the AGT was initially described as a *conceptual framework* further comprising boundaries of the theory, laws of interactions among the constructs, and the already mentioned system states trying to explain the system dynamics described by the theory in action. The predominant AGT's concern is to "deliver value" [R] to the business faster, better, and cheaper in sustainable cycles. The theory considers that a means of achieving this end, in a sustainable way, is through the combination of "agile (and/or lean) capabilities" [A] and "governance capabilities" [G] to avoid team lethargy, seeking to achieve business agility as a rational balance of sustainability and competitiveness mechanisms.

---

[3] A *unit* (or object) *of analysis* is the entity we wish to say something about at the end of the study and is considered the study's focus. The chosen unit of analysis delimits the frontier of application of the theory.

[4] *Organizational context* is an important abstraction that significantly affects Information Systems (IS) research and practice, and its effectiveness, as well different components of the organizational context constitute different environments in which IS are developed and implemented (Xu et al., 2011).

In order to interpret the AGT as a system, Luna et al. (2020) combined explanations and predictions from the theory propositions and system states into *eight[5] theoretical scenarios*[6]. They then analyzed the system created by the emerging theory and compared it to the data collected in the real world. The authors paid particular attention to the application of the scenarios, the behavior, usefulness, and consequences, considering the *organizational contexts* observed *as the object of analysis of those scenarios*. In short, the authors characterized each *theoretical scenario* developed as follows:

(1) **Beginners' Scenario ($\varphi_0$)** - in which the *organizational context* under analysis has no experience with governance [G], neither an agile and/or lean culture previously established [A].

(2) **Governance Experienced Scenario ($\varphi_1$)** - representing the organizational context in which it has developed some governance experience, considering that governance capabilities [G] can put business operations [B] under control but at the cost of excessive *bureaucracy*. Organizational contexts that fit this scenario seek to simplify and develop business agility without losing steering capability.

(3) **Agile or Lean Culture Scenario ($\varphi_2$)** - representing organizational contexts in which it has developed some agile or lean culture, and they consider that agile capabilities [A] can boost business operations [B]. However, they feel that agility without steering capability might be dangerous. Consequently, they seek to ensure that business operations [B] must be controlled, however, without losing the benefits of agile capabilities [A].

(4) **Dissociative Scenario ($\varphi_3$)** - where there is significant evidence of both *governance experience* and *agile/lean culture*, which may be an indication that there are latent agile [A] and governance [G] capabilities in the organizational context. However, these capabilities do not combine (together) to fulfill the organizational context goals.

(5) **Startup Scenario ($\varphi_4$)** - characterizes the theory startup, in which almost all *laws of interaction* from theory might be observable in full. The exception is made to Law 5 (*Sustainability and competitiveness*), which manifests itself only in scenario ($\varphi_5$) as a natural consequence of the gradual internalization of Law 1 over time.

(6) **Countermeasures Scenario ($\varphi_5$)** - it is reached by the organizational context because of the combined and coordinated work of [A] and [G], started in **scenario ($\varphi_4$)**. The *countermeasure behavior* is explained by Law 5, which can be decomposed into two theoretical *sub-scenarios*:

  i. *Sustainability* scenario ($\varphi_{5'}$) - in which agile capabilities [A] and governance capabilities [G] interact to reduce the effects of moderator factors [M] in the organizational context; and,

  ii. *Competitiveness* scenario ($\varphi_{5''}$) - where agile [A] and governance capabilities [G] collaborate to decrease the effects of environmental factors [E] upon the organizational context.

(7) **Dynamic Scenario ($\varphi_n$)** - depicts when an organizational context experiences an elevated level of organizational sustainability and competitiveness. People develop their activities in a superior level of awareness, acting and reacting in an *unconsciously competent* manner, as a coordinated whole, almost intuitively, to deal with the emerging issues from the organizational context and the environment where they are inserted. In this *scenario*, [E] and [M] have a minimal effect on the organizational context, and [A], [G], [B], and [R] have their maximum values.

Those scenarios, their interrelationship, and the evolution of the theory behavior over time are depicted in (Luna et al., 2020). APPENDIX A illustrates the 'big picture' based on (Luna et al., 2020), in which they are portrayed in order of increasing complexity over time in four lanes. We interpret the scenarios' sequence as the order of evolvement for the system described by the emerging theory, which can be used as roadmaps to help start the theory application. For instance, the **Beginners' Scenario ($\varphi_0$)** considers that the organizational context under analysis (e.g., a team) has no experience in governance, neither an agile nor lean culture previously established, so that [G] and [A] constructs do not manifest themselves in this scenario. Hence, the effects from the external environment [E] (outside of the team, e.g., from enterprise) are felt directly by business operations [B] and indirectly through the restraining performed by the effects of the moderator factors [M] of the team, in its inner context.

AGT's key constructs and core assumptions were broken down into 16 hypotheses, combined in a particular set of theoretical scenarios already characterized. **Fig. 1** represents the whole picture of the hypotheses operationalized to test the theory. Each hypothesis was designed as simply as possible to facilitate its test by *Structural Equation Modeling* (SEM). The full description of each hypothesis is characterized in APPENDIX D. For instance:

- $H_1$ states that "*agile capabilities [A] positively influence governance capabilities [G]*".

- While $H_2$ asserts that "*governance capabilities [G] positively influence business operations [B]*".

- In turn, $H_3$ declares that "*agile capabilities [A] positively influence business operations [B]*".

---

[5] We list eight scenarios because the fifth scenario, *Countermeasures Scenario ($\varphi_5$)*, occurs in two theoretical sub-scenarios.

[6] They are distinct combination of explanations and prediction based on the emerging theory, describing a sequence of characteristic events (Luna et al., 2015).

- Whereas $H_4$ describes that "*business operations [B], under the influence of agile capabilities [A] and governance capabilities [G], have a positive influence on value delivery [R]*".

- Eventually, $H_{16}$ affirms that *"governance capabilities [G] positively and partially mediate the relation between agile capabilities [A] and business operations [B]"*, and so on.

These sixteen hypotheses are focused on the theory's central propositions, and their assessment can help to establish the underlying logic upon which the theory's system states, which describe their systemic behavior, are based. An observant reader will realize that the hypothesis $H_{16}$, previously stated, is not depicted explicitly in **Fig. 1**. Indeed, $H_{16}$ is described by the *mediation* between the constructs [A] (*predictor variable*) and [B] (*dependent variable*) through [G] (*mediator variable*). Therefore, it relates to the subsystem comprised of $H_1$, $H_2$, and $H_3$: referring to the $1^{st}$ *Law* of the theory, which states: "*Agile governance arises when agile capabilities [A] are combined and coordinated with governance capabilities [G], activating or intensifying an increase in the level of business operations [B], which in turn increases value delivery [R]*".

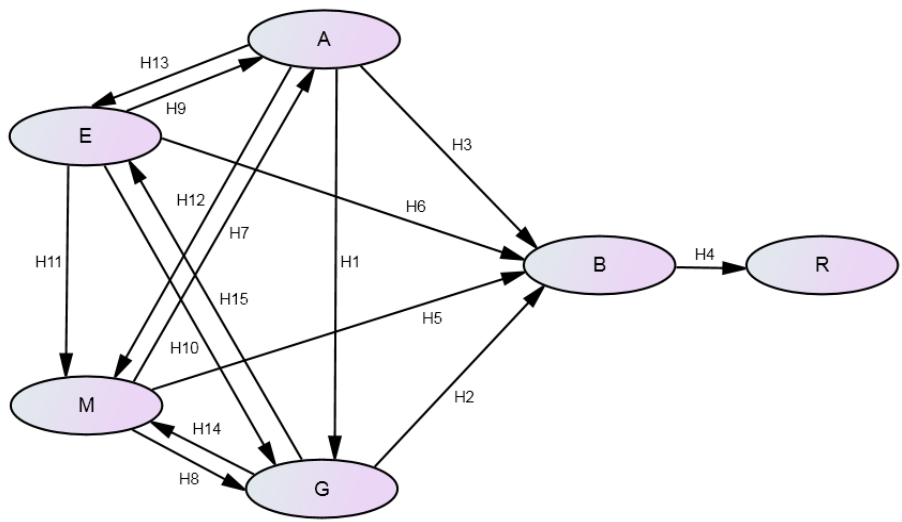

FIG. 1. **THEORY'S HYPOTHESES: VISUAL CHARACTERIZATION. ADAPTED FROM:** (Luna et al., 2020).

Considering the **Startup Scenario ($\varphi_4$)**, all the theory's constructs are present. The 5$^{th}$ law does not show up because it only manifests itself in the **Countermeasures Scenario ($\varphi_5$)** and **Dynamic Scenario ($\varphi_n$)**. Consequently, only hypotheses $H_{12}$, $H_{13}$, $H_{14}$, and $H_{15}$ are absent in this scenario. For instance, in this scenario, the *sensing ability* [A] can empower the team *steering ability* [G], improving the capacity of business operations [B] to perceive and react coordinately on time to changes, contributing to better value delivery [R] to the consumers. This behavior can be tested through hypotheses $H_1$, $H_2$, $H_3$, $H_4$, and $H_{16}$.

The *readiness dimension* [A] can be combined with the *strategic alignment* [G], developing the ability to *keep teamwork strategically aligned with the business goals*. At the same time, the team develops a mindset to be ready, willing, and able to do what is needed to achieve strategic objectives and seek to associate each routine activity with the overall strategy. Alternatively, even when the team's *positive attitude* [A] can become the *control* [G] activities *less oppressive* by developing the *ability to self-control, collaboration, and appreciative influence* to do, helping to engage team members. As a consequence, *deadline monitoring* [G] can improve the *punctuality of the deliveries* from the software development process [B] (outcomes), which in turn *reduces the time to deliver software features* at the right time to the business demands (timely delivery) [R].

The theory has been conceptualized (Luna et al., 2015) and operationalized (Luna et al., 2020) but has not yet been assessed/tested in the real world. Although some studies have tried to investigate whether the principles derived from AGT (Halim et al., 2021) are supported in practice, the test of AGT's key hypotheses and core assumptions remains unexplored in depth. In this report, we assess the AGT in an empirical study. Notwithstanding were identified *sixteen hypotheses* to represent the entire system described by the theory, **we infer that the hypotheses $H_1, H_2, H_3, H_4,$ and $H_{16}$ are the most representative** because they are based on the theory's central propositions. Furthermore, as such, they can accurately assess the theory's plausibility since they are closely related to the core behavior of the system and, therefore, to the theory's essence.

## 3. METHOD

This study focuses on assessing the AGT's key hypothesis and core assumptions in practice through an empirical study with a group of practitioners. We adopt quantitative and explanatory research through a survey (Groves et al., 2013). **Fig. 2** depicts the *research framework* developed to guide this study.

Further, **Table 1** depicts the procedures related to each step illustrated in **Fig. 2** and the Sections in this report where everyone is detailed. The procedures described in **steps 1** and **2** are related to the *Study Design*, **step 3** comprehend the *Study Execution*, **steps 4, 5**, and **6** are related to the study *Data Analysis* procedures, and **step 7** concerns the study *Data Synthesis*.

### 3.1. Study Design

We frame this study using a quantitative inquiry approach by deriving eight SEM models from the AGT theoretical scenarios. We consider theory elements of cause and effect since we not only presupposed the existence of relationships between variables in the model but also assumed the directionality of interactions, described by our hypotheses derived from them. For these reasons, an "*explanatory survey*" was considered the most appropriate (Creswell, 2003; Groves et al., 2013). Then we sought to evaluate each model separately and, in sequence, consolidate the results by considering the system described by the AGT in action, aiming to answer the research question.

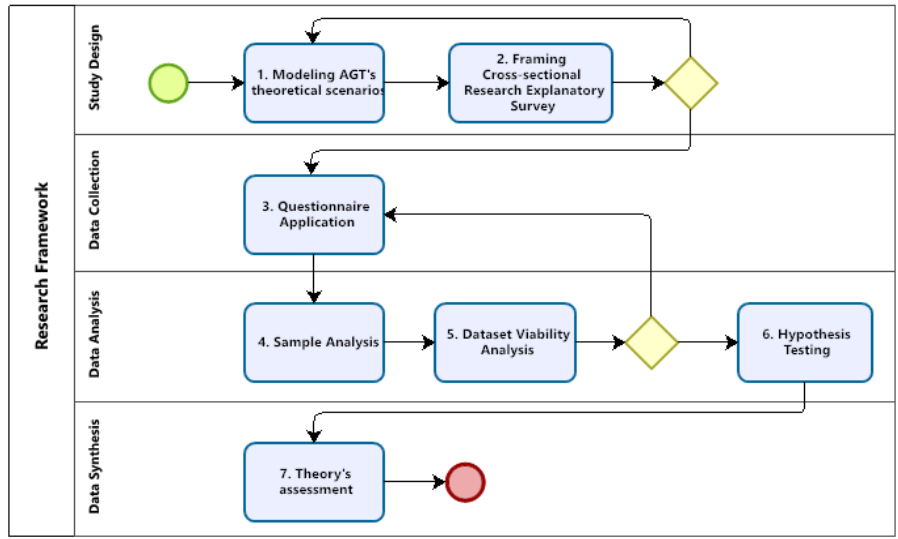

FIG. 2. **RESEARCH FRAMEWORK.**

As described in Section 2, a*gile governance phenomena* occur continuously over time. This study is "cross-sectional" because the perception of the representative agents from those phenomena is evaluated in a specific time and space, considering the period in which the survey research was conducted. Hence, we include selected evidence from a *cross-sectional research explanatory survey*.

TABLE 1. **STEPS AND PROCEDURES OF THE STATISTICAL METHOD.** INSPIRED BY: (Severo et al., 2020).

| Step | Procedures |
|---|---|
| **1. Modeling AGT theoretical scenarios** | Considering the development of eight Structural Equation Models: $\varphi_0 .. \varphi_n$ (described in Section 3.1.2). |
| **2. Framing Cross-sectional Research Explanatory Survey** | Considering: Sample characterization, Sample sizing, Survey Protocol, Questionnaire Development, and Questionnaire Pilot (Section 3.1) |
| **3. Data collection** | Considering the Questionnaire application (Section 3.2). |
| **4. Sample Analysis** | Considering:<br>• Analysis of sample adherence to the sample profile criteria (at the beginning of the Section 4).<br>• Sampling Groups Analysis and Demographic Analysis (Section 4.1).<br>• Participants' Mindset and Experience Analysis (Section 4.2). |
| **5. Analysis of the data set viability** | Considering: Independence of observations, Scale recoding and treatment of missing values, Internal consistency of the measuring instrument, Univariate and multivariate normality, Nonzero sample covariance, Multicollinearity absence, and Absence of non-standard values (outliers) (Section 4.3.1). |
| **6. Hypotheses testing** | Considering:<br>• SEM Analysis and Models' Adjustments (Section 4.3.2).<br>• Confirmatory Factor Analysis (CFA) (Section 4.3.3).<br>• Validity related to the constructs (Section 4.3.4).<br>• Model Estimation for validity analysis (Section 4.3.5).<br>• Path Analysis (Section 4.4). |
| **7. Theory assessment** | Considering:<br>• Analysis of the AGT's hypotheses (Section 5.1).<br>• Findings interpretation (Section 5.2).<br>• Strength of evidence (Section 5.3).<br>• Construct validity (Section 5.4.1).<br>• Internal validity (Section 5.4.2).<br>• External validity (Section 5.4.3). |

We applied *Structural Equation Modeling* (SEM) and *Confirmatory Factor Analysis* (CFA) (Hooper et al., 2008; Marôco, 2014; Weston, 2006) to theory assessment. Marôco (2014) highlights SEM as a generalized modeling technique used to test the validity of theoretical models, which define hypothetical and causal relations among variables. Using SEM, the researcher starts by formulating the *theoretical framework* (model) and then collect data to confirm, or not, this theoretical framework. The *theory is the engine of analysis*, contrary to the *classical statistical paradigm* in which data, not the theory, is at the heart of the research process. CFA allows us to test the hypothesis that a relationship between the observed variables and their underlying latent construct(s) exists in each of the eight theoretical scenarios developed by Luna et al. (2020) and characterized in Section 2.

### 3.1.1. Modeling AGT theoretical scenarios

As mentioned in **Section 2**, the AGT's key constructs and core assumptions were broken down into 16 hypotheses (depicted in **Fig. 1**), combined in a particular set of *theoretical scenarios*, generating eight theoretical models derived from AGT operationalization. Those theoretical scenarios were assessed by "representative agents of the phenomena under study", which we will name *practitioners*. Their responses were analyzed using *Structural Equation Modeling* (SEM) (Hooper et al., 2008; Marôco, 2014; Weston, 2006).

We tested the theory in multiple scenarios because AGT advocates the dynamic existence of these eight scenarios, which can manifest depending on the level of awareness the organizational context has regarding its experience with the application of agile governance. The reader should understand each AGT theoretical scenario as a frame (a static photo) of a dynamic sequence of events (a video), which portrays situations experienced in developing capabilities necessary for the governance of business agility. The AGT's system states describe the complete sequence of these events (Luna et al., 2020). However, this 'movie' can occur differently in each organizational context analyzed by the AGT. Depending on the scenario characteristics and the moment experienced by each organization, some constructs and laws of interaction might be present or not affecting the unfolding of observed events. APPENDIX A depicts the different possible paths for the evolution of the theoretical scenarios of AGT, instrumentalizing the analysis and description of real scenarios experienced by organizations. Assessing all those eight scenarios means teste the whole AGT breadth of application.

Despite were identified *sixteen hypotheses* to represent the entire system described by the theory, in keeping with Luna et al. (2020), we **infer that the hypotheses $H_1$, $H_2$, $H_3$, $H_4$, and $H_{16}$ are the most representative, and as such, can accurately assess the plausibility of the theory since they are closely related to the core behavior of the system and, therefore, with the theory's essence**.

### 3.1.2. Models derived from the theoretical scenarios of AGT

To formulate the models for each AGT's theoretical scenario, we proceed as follows. Firstly, we represent the *structural sub-model* as a *paths diagram* showing the causal relationships among the constructs manifested in each scenario. Secondly, the hypotheses (see APPENDIX D) present in each scenario were depicted in the structural sub-model. Thirdly, the observed variables (*empirical indicators*[7] identified in APPENDIX B) to measure the *expression* of every construct existing in the scenario are associated with the respective *construct*, forming the *measure sub-model*. Finally, the *complete structural equation model* was presented, with its structural component and measurement.

For instance, **Fig. 3** depicts the SEM model designed for the AGT **Dynamic scenario ($\varphi_n$)**. We have chosen this scenario to illustrate the development of the theoretical models because it is the most complex theoretical scenario described by the AGT, i.e., where all the constructs and their relations (hypotheses) are present. However, each theoretical scenario has been modeled and tested independently, generating its specific model.

Every model consists of two sub-models: i) **structural sub-model**: illustrated by the gray area in **Fig. 3**; and ii) **measurement of sub-model**: the set of indicators outside the gray area. In **Fig. 3**, the structural sub-model includes *six constructs* (ovals) and their interactions or causal relationships (one-way arrows). The measurement of the sub-model is formed by particular sets of *manifest variables (rectangles)*. The manifest variables correspond to an empirical indicator (see APPENDIX B), depending on the constructs present in that scenario.

Describing the causal model depicted in **Fig. 3**, we can observe **six** *endogenous* (dependent) *variables*: **Effects of environmental factors [E] ($\eta_1$)**, operationalized by **five** *dependent* (manifest) *variables* ($y_1$ to $y_5$); **Effects of moderator factors [M] ($\eta_2$)** whose set of measurement is formed **five** *indicators* ($y_6$ to $y_{10}$); **Agile capabilities [A] ($\eta_3$)**, operationalized by **four** *dependent variables* ($y_{11}$ to $y_{14}$); **Governance capabilities [G] ($\eta_4$)**, measured by **four** *manifest variables* ($y_{15}$ to $y_{18}$); **Business operations [B] ($\eta_5$)**, operationalized by **three** *dependent variables* ($y_{19}$ to $y_{21}$); and, **Value delivery [R] ($\eta_6$)** measured by **three** *manifest variables* ($y_{22}$ to $y_{24}$).

The error terms of the independent (manifest) variables are represented by δ. The error associated with each dependent variable is represented by ε ($\varepsilon_1$ to $\varepsilon_{24}$). Both δ and ε represent the part not explained by the factor (construct) the variable manifest and which would be explained thus by other factors (variables) not considered in the model. Similarly, endogenous latent (dependent) variables, $\eta_1$ to $\eta_6$, have their causes in their relations, and the unexplained part is attributable to the **error** or disturbance $z$ ($z_1$ to $z_6$).

---

[7] An empirical indicator is a concrete and specific real-world proxy used to develop and assess a middle or a situation-specific theory. The function of empirical indicators is to provide how theories are generated or tested by measuring the theory constructs, considering that sometimes it is not feasible to measure constructs directly (Fawcett, 2021).

The factorial weights, represented in the model by the Greek letter *lambda* ($\lambda_{y_1}$ to $\lambda_{y_{24}}$), and the structural coefficients characterized by the hypotheses ($H_1$ to $H_{15}$), are depicted in the order of cause and effect. For example, considering **Fig. 3**, based on the theoretical scenario $\varphi_n$, the factorial weight of the factor $\eta_2$ [M] in $y_6$ [*Organizational culture refractoriness*] is $\lambda_{y_6}$.

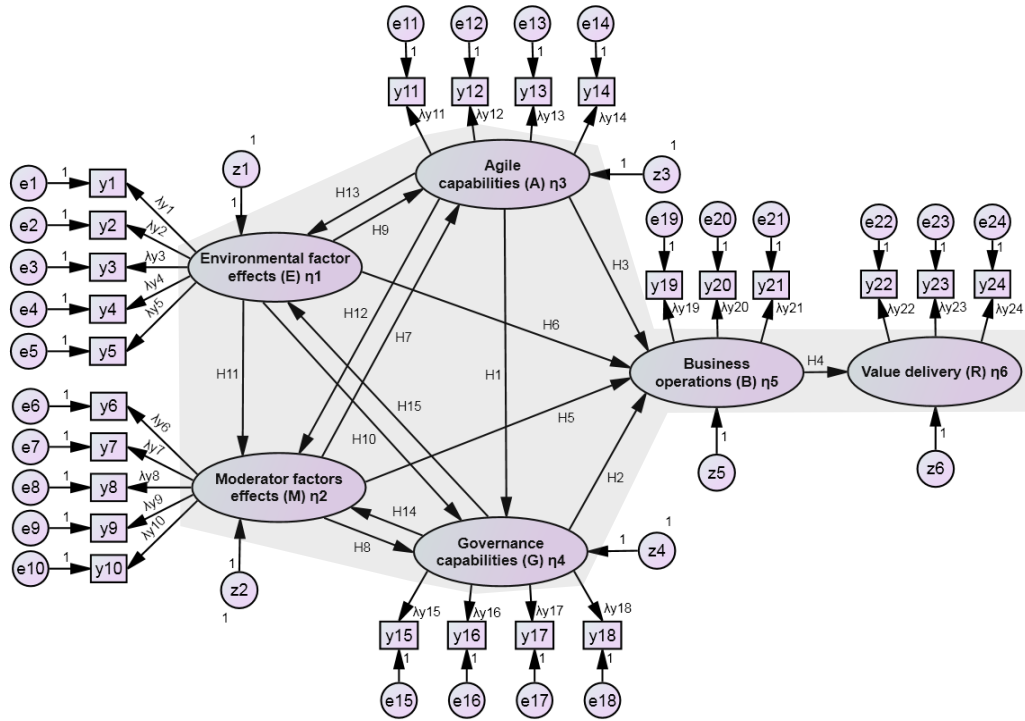

FIG. 3. **SEM Theoretical model based on the Dynamic scenario ($\varphi_n$)**.

We elaborate on the SEM model (**Fig. 3**) to be as similar as possible to the Theory's Hypotheses diagram depicted in **Fig. 1**. Correspondingly, $H_1$ is the structural coefficient (or regression coefficient) between $\eta_3$ and $\eta_4$ (respectively, constructs [A] and [G]), as well as this coefficient, represents the homonymous theory's hypothesis ($H_1$) on the model. **Fig. 3** depicts one of the eight theoretical research models designed for this study, with the hypotheses that foresee the influence relations between the constructs of the Agile Governance Theory (AGT). The remaining research models derived from the other theoretical scenarios are depicted in APPENDIX E.

### 3.1.3. Equations derived from the theoretical scenarios of AGT

A set of linear equations can alternatively represent the SEM model of **Fig. 3**. For clarity, we have replaced *Greek symbols* relating to every construct with the representative letters of each theory's units using the *modern English alphabet*. As a result, **Equation F.7** in APPENDIX F is the *General Equation of the Structural Theoretical Model*, derived from the most sophisticated theoretical scenario depicted from Agile Governance Theory, **Dynamic scenario ($\varphi_n$)**, can be transcribed in colloquial terms as:

> "*Value delivery [R] can be described as the result of the influence of the disturbing Effects of external environmental factors [E], and the restraining Effects of inner moderator factors [M]; as well as the enhancers effects from Agile capabilities [A] and Governance capabilities [G], upon the Business operations [B], and their interactions, into the organizational context under analysis*".

TABLE 2. **Profile: Representative agents from Agile Governance phenomena.**

| Criteria | Description |
| --- | --- |
| (C1) Role | Researcher (scholar) and/or Practitioner. |
| (C2) Experience in the topic domain | (a) **Governance capabilities**[8], and (b) **Agile capabilities**[9]. |
| (C3) Responsibility level | Considering the context of leadership, coordination, management, or direction. |
| (C4) Experience in | Worked in: a team, project, business unit, enterprise, or multi-organizational setting. |

---

[8] It is the ability to develop competencies related to how an organizational context is conducted, administered, or controlled, i.e., strategic alignment, decision-making, control, compliance ability, steering skills, policymaking, and accountability, among others.

[9] It is the ability to develop competencies related to principles, values and practices, from agile, and lean philosophy, i.e., flexibility, "doing more with less", agility, adaptability, resilience, responsiveness, "coordinability" or "orchestrability", self-organization, simplicity, readiness, among others.

### 3.1.4. Framing Cross-sectional Research Explanatory Survey

The models are tested against the obtained measurement data to determine how well they fit the data (Pearl, 2000). Due to the nature of the phenomena under study and the multidisciplinary nature of agile governance, we use the *theory's external boundary-determining criteria*, characterized in (Luna et al., 2020), to define the sample and profile of the participants as depicted in **Table 2**.

The sample is non-probabilistic for convenience (Hair et al., 2010). The *Explanatory Survey* was administered to 956 representative agents of the agile governance phenomena, considering the profile depicted in **Table 2**. They were organized in distinct sampling subgroups, considering experts and researchers found from the systematic review published by Luna et al. (2014), practitioners and scholars interviewed during AGT development, and practitioners from Professional Groups based on Social Networks identified in (Luna et al., 2015).

As a result, 281 respondents were obtained, which meets the sample size requirements of more than 100 valid cases, according to Westland (2010). This author suggests the use of **EQUATION 1**, in which (p) means the *number of items or manifest variables* and (f) is the *number of latent variables* (or factors) of the model, as well as (r = p / f) determines the *number of manifest variables by construct*.

$$n_{SEM} \geq 50r^2 - 450r + 1100$$   **EQUATION 1. SEM SAMPLE SIZING.** SOURCE: (Westland, 2010).

According to **EQUATION 1**, the *SEM and CFA Analysis* for the model of **Fig. 3**, related to the most complex scenario of AGT: **Dynamic Scenario ($\varphi_n$)**, comprising six constructs and 24 manifest variables, in which r = 4, would require at least 100 observations (valid cases).

$$n_{SEM} \geq 50 \times 4^2 - 450 \times 4 + 1100, \therefore n_{SEM} \geq 100$$

From the 281 responses collected, which gives a response rate of around 29.4%, 163 cases were discarded because they did not fully meet the criteria established for research (such as not fully completing the survey questions), resulting in 118 valid cases (n = 118), which represent a global effective response rate of over 12.3%. The resulting useful sample for statistical analysis, with meaningful geographical coverage, was composed of 118 respondents from 86 distinct organizations and 19 countries, representing a statistically significant sample since the minimum sample size must have at least 100 cases, according to the chosen assessment methods.

We developed a research protocol for the *cross-sectional explanatory survey* by adapting several authors' policies, procedures, and techniques. Grossman et al. (2009) assisted with the survey guidelines. Walonick (2012) helped us to test the survey's reliability. Passmore et al. (2002) aided us in developing the data-collecting instrument. Fowler (2009) and Kitchenham et al. (2002) guided us in choosing the level of rigor appropriate to the study. We also consult specialists on the topic and methods adopted. This protocol was available on the web for consultation by all participants during the study.

The instrument of data collection (questionnaire) was developed based on the broader scenario for theory instantiation leading us to the **Dynamic Scenario ($\varphi_n$)**. The questionnaire[10] presents 41 questions, four related to the participant's mindset, eight associated with the profile of the respondents, five about the study analysis, and 24 addressed to theory assessment, worded as statements and organized into the AGT constructs: *Environmental factor effects[E], Effects of moderator factors [M], Agile capabilities [A], Governance capabilities [G], Business operations [B],* and *Value delivery [R]*. The questionnaire consists of statements in which the respondent chooses an alternative answer on a 10-point Likert scale (from 1 totally disagree to 10 totally agree). We adopted an ordinal scale of 10 points by following the recommendations of (Lomax and Schumacker, 2012) and (Marôco, 2014) because of the statistical method chosen by this study: *Structural Equation Modeling* (SEM). The questions from AGT Survey are listed in APPENDIX G.

Following the suggestions of Pinsonneault and Kraemer (1993), the questionnaire was previously tested in form and content, along with 12 scholars and practitioners members of the sample population, employing semi-structured interviews, using the questionnaire as a script. In general, the *pilot* questionnaire evaluated the clarity, validity, reliability, and relevance of the questions in statistical terms, in the "agreement spectrum" from the measurement scale. Some comments, suggestions, and considerations about the questionnaire's specific points helped refine and improve it.

### 3.2. Data Collection

The questionnaire was applied online, using SurveyMonkey[11], organizing the 956 invited participants (sample) in 18 distinct sampling subgroups. These sampling groups were classified based on "interest groups"[12] and stratified by language, which proved especially useful when planning and implementing the communication plan for each group. The *communication plan* for every sample group comprised: an invitation email, a reminder email, and the last reminder email. Eventually, when a respondent had any doubt about the questionnaire, we kept in touch individually by email or, where required, instant messaging tools (e.g., Skype, Google Hangouts/Meet, WhatsApp, among others).

---

[10] The final version of the questionnaire (preview) is available at http://bit.ly/agt_survey
[11] SurveyMonkey is an online survey software available at https://surveymonkey.com/
[12] For instance: Authors and Researchers on Agile Governance, Experts, Scholars, and Practitioners from Professional Groups based on Social Networks.

The research also used the snowball method. The researchers sent the questionnaire to their contacts and received suggestions from other potential participants. Thus, the profile of the subject suggested was verified if has matched the criteria of **Table 2**. Consequently, the subject was accepted or rejected by the research team. When accepted, they received a personalized invitation.

Our sampling composition adopted a *mix of purposive sampling* types (Patton, 1990). This approach starts with a purpose in mind, and the sample is thus selected to include people of interest and excludes those who do not suit the purpose. Participants are chosen because of some characteristics. In our case, according to the *sample profile* depicted in **Table 2**. Details from the *sample analysis* will be discussed in Section 4.1.

TABLE 3. **QUALITY INDICES OF THE SEM MODEL ADJUSTMENT.** SOURCE: (Marôco, 2014) AND (Mulaik, 2009).

| Index | Variable | Reference values |
|---|---|---|
| $\chi^2$; p-value | It measures the discrepancy between the theoretical model and the data sample. | The smaller, the better; p > 0.05 |
| $\chi^2/gl$ | It being the sensitive chi-square to the sample size, it is useful to standardize the index by dividing it by the degrees of freedom. | > 5; bad fit<br>] 2, 5]; acceptable fit<br>] 1, 2]; good fit<br>~ 1; Very good fit |
| **CFI** (*Comparative Fit Index*)<br>**GFI** (*Goodness of Fit Index*)<br>**NFI** (*Normed Fit Index*)<br>**TLI** (*Tucker Lewis Index*) | It standardized incremental indices that measure the model fit for a specific range of values. | <0.8; bad fit<br>[0.8; 0.9 [; acceptable fit<br>[0.9; 0.95 [; good fit<br>≥0,95; very good fit |
| **AGFI** (Adjusted Goodness of Fit Index) | GFI adjusted for degrees of freedom of the model. | > 0.9 |
| **RMSEA** (Root Mean Square Errors of Approximation) | It measures the quality of the model that fits the covariance matrix of the sample, considering the degrees of freedom. | > 0.10; unacceptable fit<br>] 0.05, 0.10]; good fit<br>≤0.05; very good fit<br>p-value ≥0.05 |
| **Parsimony Indices:**<br>• *Parsimony GFI (PGFI), based on GFI*<br>• *Parsimony CFI (PCFI), based on CFI*<br>• *Parsimony NFI (PNFI), based on NFI* | These relative fit indices adjust to most of the above ones. The adjustments are to penalize less parsimonious models, so simpler theoretical processes are favored over more complex ones. Mulaik (2009) developed a number of these. Although many researchers believe that parsimony adjustments are important, there is some debate about whether they are appropriate. Many authors agree that researchers should evaluate model fit independent of parsimony considerations. However, there are alternative theories favoring parsimony. With that approach, we would not penalize models for having more parameters, but if simpler alternative models seem good, we might favor the simpler model. | When the more complex the model, the lower the fit index.<br>Most scholars widely accept values for those indices >0.60 in this topic. |

### 3.3. Overview of Data Analysis and Synthesis

From the sample obtained by applying the questionnaire, the *data analysis* was performed based on **steps 4**, **5**, and **6** from the *research framework* illustrated in **Fig. 2** and the respective statistical analysis procedures depicted in **Table 1**. For analyzing model estimation results, in **step 6**, we used the SEM quality indices for the model fit described by Marôco (2014) (p. 55) and Mulaik (2009), depicted in **Table 3**.

In turn, the *data synthesis* was performed based on **Step 7** from the *research framework* illustrated in **Fig. 2** and the respective statistical analysis procedures depicted in **Table 1**, which were required to assess the study hypotheses and interpret the results. We have used software IBM® SPSS® Statistics 20.0 and IBM® SPSS® Amos 20.0. This approach related to the meaning evaluation of every hypothesis test and its consequent signification for the theory was an essential instrument for convergence and synthesis for assessing the AGT. Further information about *data synthesis* is discussed in Sections 4 and 5.

Models, data, and other files that support research findings are available in the research repository.

## 4. RESULTS

A total of 118 individuals adherence to the sample profile criteria, based on Table 2, concerning **role [C1]**, 118 of the subjects (100%) were practitioners: 46 of them (39%) were researchers (scholars), and former practitioners; 72 of them (61%) were currently active practitioners. Regarding **experience in the topic domain [C2]**, all the respondents were experienced. Thirty-two participants (27.1%) have up to 10 years of work experience (Q29); 38 respondents (32.2%) have between 10 and 20 years of work experience; and 48 (40.7%) have more than 20 years. In addition, participants (100%) had experience in both governance and agile/lean, highlighting: that 64 of them (54.2%) have more than six years of "Governance experience" (Q32), and 44 of them (37.3%) have more than six years of "Agile/Lean experience" (Q33). Respecting **responsibility level [C3]**, all participants worked in leadership positions: 16 (13.6%) have distinguished leadership "Job positions" (Q30) in their organizations, as CEO, CIO, or Business Owner; 54 (45.8%) have leadership "Job positions" (Q30) in their organizations, as Executive, Public Administrator or Project Manager; and 48 (40.7%) worked as coordinators, managers, or team leaders. Finally, relating to **experience in [C4]**, all participants had worked in the organizational contexts described by AGT, as follows: 19 (16.1%) teamwork; 32 (27.1%) project; 22 (18.6%) business unit; 34 (28.8%) enterprise; and 11 (9.3%) multi-organizational setting.

Besides, the number of valid cases for analysis overcame the minimum sample size calculations set by EQUATION 1, and our sample achieved the required statistical significance, as described in Section 3.1.

### 4.1. Sample Analysis

The most representative sampling groups were "Industry" (52, 44.1%), which comprises practitioners according to the required *respondent profile*. It was followed by "Academy" (31, 26.3%), which holds scholars and researchers; "Government" (20, 16.9%) which consist of practitioners who develop their activities in the public administration context; and "Agile governance Researchers and Authors" (15, 12.7%), which encompasses the "Authors" found as a result of SLR-AG.

We may also analyze the sample by the purposive sampling types, according to Patton (1990). The most representative sample type was "**Typical cases**" (89, 75.4%), which allows us to understand the responses profile obtained as "*what is agreed as average, or normal*", arising out of the representative agents from the phenomena in the study. This sample representativeness means we can compare the findings from this study using typical case sampling with other similar samples (i.e., comparing samples but not generalizing a sample to a population).

Still, in keeping with Patton (1990), while typical case sampling can be used exclusively, it may also follow another type of purposive sampling technique, such as "**Snowball or chain sampling**" (14, 11.9%) and "**Critical cases**" (15, 12.7%). This prior one is our study's second most representative purposive sampling type. The "Snowball" sampling also went through the process of verifying whether each respondent, who was suggested (candidate) by any participant invited/selected to participate in this study, has matched with the same respondent profile adopted to choose the "Typical Cases" (see **Table 1**). This approach implies the "Snowball" sampling as a "Typical case". Hence, we can infer that the "Typical cases" represent around 87.3% of the purposive sampling types found in our survey study.

The "**Critical cases**" was this study's third most representative sampling type. Although sampling for one or more critical cases may not yield generalizable findings, they may allow us to develop logical generalizations from the evidence produced when studying a few cases in depth. However, the representativeness of this sampling type is too small to allow any logical generalization from the findings. This latter group comprises the top experts in the phenomena under study and represents 12.7% of our survey sample.

Concerning the geographical location of each respondent[13], we can identify the sample share by country and, consequently, by continent, comprising 86 organizations in 19 countries. Forty-seven (39.8%) participants have located abroad from Brazil (71, 60.2%), in distinct countries covering five continents. The second most represented country was Canada (10, 8.5%), followed by the US (9, 7.6%); France (4, 3.4%); UK and Netherlands both with (3, 2.5%); Germany, Malaysia, New Zealand, Switzerland, and Slovenia (2, 1.7%); and Cameroon, China, Croatia, Finland, Italy, Slovakia, Spain, and Venezuela each one with (1, 0.8%). This information gives us an overview of the geographical coverage of the study.

Only 19 subjects (16.1%) chose not to identify their companies, allowing us to count participants from 86 companies in our study. The study's participants mostly (55.9%) work in large "Organization size" (Q34), which has a multinational or global "Operating scale" (Q36) in 24.6% of the cases, on a broad spectrum of "Economy sectors" (Q35), from which stands out *IT industry* (42.4%), some of them with worldwide presence such as IBM, Huawei, Accenture, Furukawa, among others. Besides, we identified the participation of *scholars from renowned institutions* such as University College London, University of British Columbia, Delft University of Technology, Auckland University of Technology, and Villanova University, to name a few. We also recognized organizations related to *Public Administration*, such as British Columbia Housing (Canada); the Brazilian Ministry of Planning, Budget, and Management; State Departments (Education, Healthcare, Treasury, Comptroller); and Courts (Justice, Auditor's Office, Attorney's Office) of some Brazilian provinces; DATAPREV; INFRAERO, for instance. Also, other *Services and Industries* were represented by companies such as LATAM Airlines, Petrobras, Brasbunker/ Bravante, Susquehanna International Group LLP (SIG), Brennand Groups in Brazil, and others.

---

[13] This analysis was based on the intersection of information provided by the respondent with the Internet Protocol (IP) automatically collected during the questionnaire fulfillment.

*4.2. Participants' mindset and experience*

Mindset is a habitual or characteristic mental attitude that determines how people will interpret and respond to situations, i.e., a complex mental state involving beliefs, feelings, values, and dispositions to act in specific ways. A closer look at these aspects is necessary once the theory moves into a practical environmental setting. The environment will influence social, economic, cultural, technological, and political factors. In contrast, positions and roles will also influence individuals of a group in the scope of an institution where an organizational context can be, in increasing order of complexity: team, project, business unit, enterprise, or a multi-organizational setting.

In our survey, two questions were related directly to the respondent's mindset: (i) what the *organizational context*[14] is chosen as a background to answer the questionnaire (Q2); and (ii) a self-assessment about how the chosen context could be evaluated in terms of **chaos and order** (Q4). The "chaos and order scale" was developed to position the agile governance phenomena. In this scale, the extreme chaos is characterized as "*chamos*", or destructive and fragmented disorder, followed by a chaos interval, a chaordic range where chaos and order coexist in a balanced way, an order area, and finally, a region of extreme control and rigidity that can become oppressive and stifling. The most representative organizational contexts (Q2) chosen were: "Enterprise" (28.8%) and "Project" (27.1%), followed closely by "Business unit" (18.6%) and "Teamwork" (16.1%). The last representative context was "Multiorganizational" (9.3%).

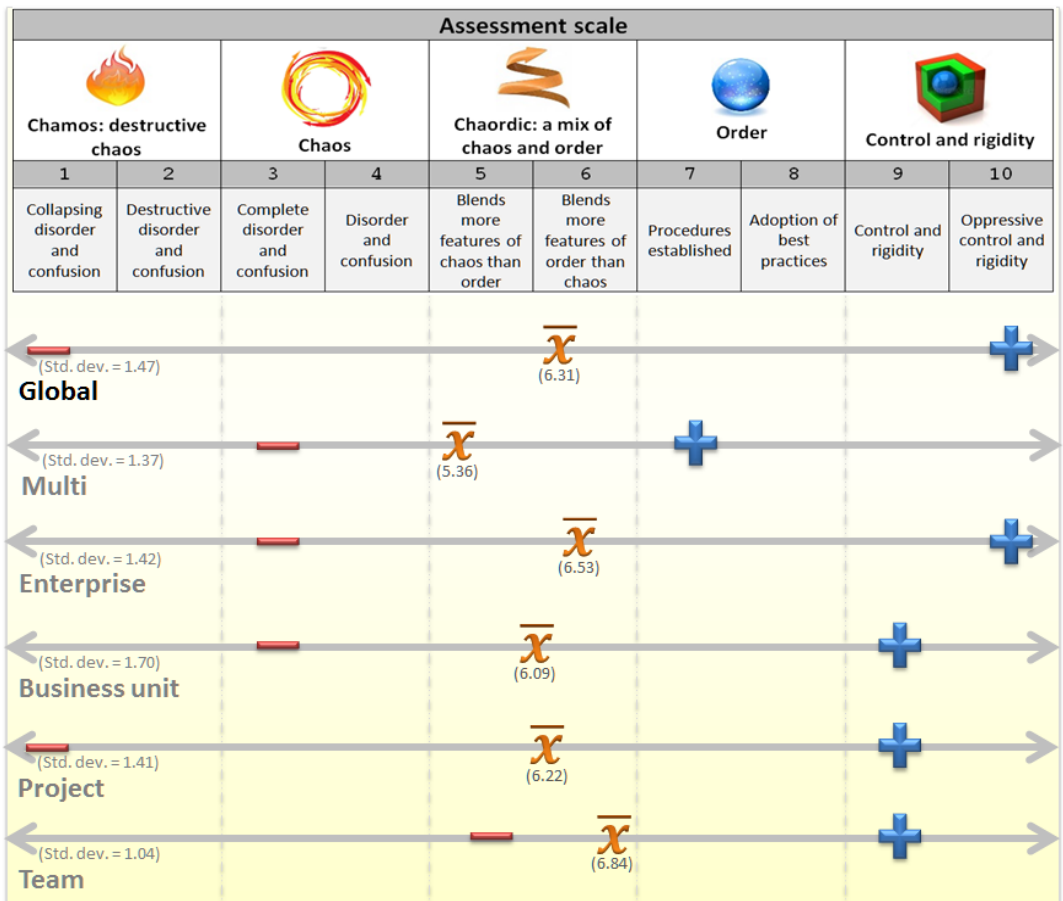

FIG. 4. CHAOS & ORDER: BY ORGANIZATIONAL CONTEXT (N = 118).

When we connect the answers from Q2 with the data gathered from Q4, we have an emerging pattern of the mindset related to chaos and order from each type of organizational context associated with this survey. **Fig. 4** depicts this cross-tabulated information between Q2 and Q4 data, presenting the *minimum value* (-), the *maximum value* (+), and the *mean* (x̄) for each context analysis. Based on the analysis of **Fig. 4**, we can infer that the organizational contexts used as the reference to answer the questionnaire have a broad spectrum of classification according to chaos and order. Indeed, the global analysis has a mean value of around 6.31 on the scale, i.e., into a *chaordic* range, corroborating the findings of the systematic review (SLR-AG) (Luna et al., 2014) and the second foundational premise of the AGT, pointed out in (Luna et al., 2015). Despite this, we cannot yet generalize this finding (see Sections 4.1 and 5.4). The results support the plausibility of the theory's assumptions, as well as highlight a symptom that needs to be further studied.

---

[14] In (Q2), we ask the subjects to keep in mind only a unique organizational context when they were answering the questionnaire, aiming to help them to establish a referential to interpret and answer each question, to minimize any misunderstanding, as well as seeking to avoid biases during our data analysis.

"Project" was the most prevalent organizational context surveyed, having results between 1 and 9 on the scale and a mean of over 6.22. This result probably indicates a broad nature (heterogeneity) of distinct projects considered to answer this survey. The results suggest that the *Enterprise* context (mean value of 6.53 and a maximum value of about 10) is a more rigid environment and less receptive to changes than a *Business Unit* context (mean value of around 6.09 and a maximum value of 9).

Another aspect that will likely influence how the agent views the theory is related to *governance and lean/agile experiences*. As described at the beginning of this section, we analyzed the survey sample regarding their experience. The experience time identified in this sampling is relevant (and consistent) enough to address aspects of agility in governance matters as a growing field of study. In other words, we can infer that these professionals have the knowledge and experience required to adequately inform about the various characteristics measured in the survey.

The dimensions explored in this Section constitute variables to assess the quality of the answers collected through the survey instrument: (1) participant mindset; and (2) participant experience. An understanding of these factors assists in the analysis that follows.

## 4.3. Data Analysis

This section presents the statistical analysis procedures required to assess the viability of the data set collected and the study hypotheses. In short, as already described in Section 3.1.1, we formulated eight theoretical models, starting from the eight theoretical scenarios from the AGT (Luna et al., 2020), aiming to assess the hypotheses related to the AGT's key constructs and core assumptions ($H_1$, $H_2$, $H_3$, $H_4$, and $H_{16}$) through the application of SEM, considering the sixteen hypotheses developed, in distinct combinations, according to each scenario.

TABLE 4. INTERNAL CONSISTENCY OF THE MEASURING INSTRUMENT: RELIABILITY AND VALIDITY, SEE SECTION 4.3.2. (N=118).

| Measurement instrument | Nº Items Before | Nº Items After | Cronbach's α (*standardized α*) Before | Cronbach's α (*standardized α*) After | Reliability analysis |
|---|---|---|---|---|---|
| Global | 24 | 20 | 0.914 *(0.915)* | 0.902 *(0.905)* | Excellent |
| Effects of environmental factors [E] | 5 | 3 | 0.714 *(0.715)* | 0.725 *(0.725)* | Good |
| Effects of moderator factors [M] | 5 | 4 | 0.861 *(0.861)* | 0.860 *(0.862)* | Very good |
| Agile capabilities [A] | 4 | 4 | 0.880 *(0.883)* | 0.880 *(0.883)* | Very good |
| Governance capabilities [G] | 4 | 3 | 0.823 *(0.835)* | 0.864 *(0.868)* | Very good |
| Business operations [B] | 3 | 3 | 0.786 *(0.785)* | 0.786 *(0.785)* | Good |
| Value delivery [R] | 3 | 3 | 0.808 *(0.809)* | 0.808 *(0.809)* | Very good |

### 4.3.1. Assumptions Verification for application of multivariate analysis

Before using multivariate techniques, which make up the structural modeling equation for the treatment of research data, we proceeded with the verification of a set of assumptions, as any transgression would seriously undermine the results and conclusions of the analysis. The following related assumptions must be verified: (1) Independence of observations; (2) Scale recoding and treatment of missing values; (3) Internal consistency of the measuring instrument; (4) Univariate and multivariate normality; (5) Nonzero sample covariance; (6) Multicollinearity absence; and (7) Absence of non-standard values (*outliers*). Every assumption was supported, allowing applying of *Structural Equation Modeling (SEM)* Analysis to the data of this research.

For instance, **(3) the reliability and validity of the measuring instrument** were evaluated by Cronbach's Alpha coefficient ($0 \leq \alpha \leq 1$), which measures the internal consistency of the data, i.e., the coherence of the answers that evaluate a given variable. As depicted in **Table 4**, we can observe that **Cronbach's Alpha is more significant than 0.7 in all cases**, suggesting good internal consistency of the questionnaire data to measure the variables under study, according to Cronbach and Meeh (1955), and Gliem and Gliem (2003).

To assess (4) **univariate and multivariate normality**, we adopted the model variation based on the *Startup scenario* ($\varphi_4$), depicted in **Fig. 5,** as a simplification of the model from **Fig. 3** because *Dynamic scenario* ($\varphi_n$) does not have *exogenous variables* (independent) necessary for this type of model estimation. In complement, ($\varphi_4$) is the complete scenario in which all constructs and the key hypotheses manifest, where the model has independent variables, and it is the model considered for the start of applying the theory.

In turn, we verify univariate and multivariate normality by applying the *Maximum Likelihood* method and observing the measures of the distribution forms, i.e., *univariate skewness* (sk), *flattening or kurtosis* (ku), and the *critical ratio* (CR). When we analyze the values of critical ratios for *univariate skewness* ($sk$) and *kurtosis* ($ku$), the absence of the violation is related to the normality assumption. However, the critical ratio for kurtosis ($CR_{ku}$) indicates the absence of normality in two variables of the model from the sample data: $y_7$ (Leanness) and $y_{11}$ (Decision making). According to Lomax & Schumacker (2012), the non-normality problem occurs because of the variable scale (ordinal rather than interval) or limited sample size. Both situations are present in this research. To solve this situation, we have followed the recommendations from Lomax & Schumacker (2012), Kline (2011), and Marôco (2014) and adopted the Bootstrap[15] estimation with 2,000 resamplings. Therefore, considering a sufficiently high number of samples, the parameter estimates do not suffer from the limitations imposed on data with a multivariate normal distribution.

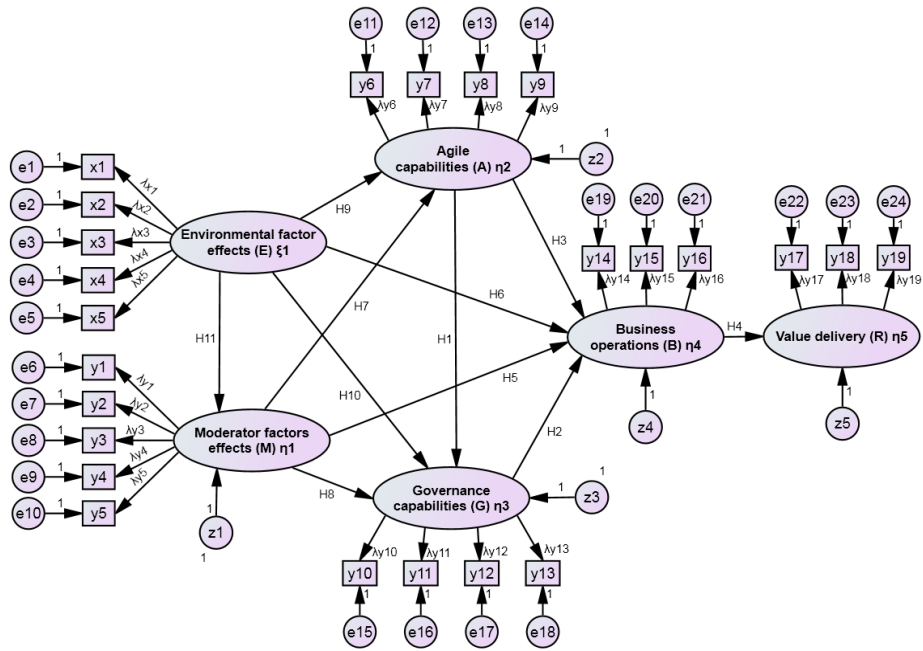

FIG. 5. **SEM THEORETICAL MODEL BASED ON STARTUP SCENARIO ($\varphi_4$).**

### 4.3.2. SEM Analysis and Models' Adjustments

Then, every model was assessed using the software IBM® SPSS® Statistics 20.0 and IBM® SPSS® AMOS 20.0 and applying the *Maximum Likelihood* method. In the model adjustment, we adopted a "two-step" approach: (1) first, adjusting the measurement sub-model; and then (2) adjusting the structural sub-model, considering the causal relations between the latent variables. Every procedure was applied to all eight SEM models derived from the AGT theoretical scenarios. As a matter of simplicity, to describe those procedures, we will continue using the model depicted in **Fig. 5**, based on the **Startup scenario ($\varphi_4$).**

In the check of the models' quality adjustment, we adopted the indexes CFI, GFI, and PCFI, having observed a good or acceptable adjustment of the values of these indices for each model according to the APPENDIX J, which depicts the goodness fit indices of the SEM model for each scenario. We also adopted RMSEA, which measures the quality of the model fit to the covariance matrix of the sample, considering the degrees of freedom. RMSEA indicates a very good fit for two and a good fit for most of the models.

To refine the models, we adopt the strategy of analyzing the *Modification Indices* (MI) based on the CFA results from the data processed (Marôco, 2014). We considered $MI > 11$ ($p < 0.001$), indicating problems with local adjustments. After evaluating the theoretical plausibility of the changes in local adjustments, the measurement errors were correlated, leading to a considerable improvement in the adjustment of the measurement model.

For instance, we have analyzed the anti-image correlation matrix to verify the adequacy of the sample measured for each variable. During the analysis, four manifest variables were eliminated ($x_2$, $x_3$, $y_2$, and $y_{13}$) due to reduced commonalities. All the adjustments performed were adequately supported by theory, related works, and evidence gathered during the development of this research. As an illustration for this example, $x_2$ has presented commonality (0.469) below 0.5. The variable $x_2$ was considered in the model to measure the influence of the "regulatory institutions" upon the organizational context and to understand the *Effects of environmental factors* [E]. Based on the comments received in question (Q7) of the questionnaire, we can imply this result has a "perception effect" related to the *respondent's mindset* regarding the *organizational context*. In this sense, people who work in enterprise and multi-organizational contexts can easily realize the effect caused by "government role" or "regulatory agencies". Whereas people who work in teams, projects, or even business units have more difficulty in seeing the "quality assurance role" or "systems auditing" as an effect of "regulatory institutions/agents". As a learned lesson, these findings strongly suggest that we must develop different questionnaires, adopting a specific group of empirical indicators (contextualized) for each organizational context and elaborating specific questions (and wording) in future studies, seeking to avoid this type of bias.

---

[15] This method provides a way to assess the empirical distribution increasing the precision of the estimates of the parameters.

The model re-specification process changed the correlation structure between the variables. For this reason, we re-assessed the reliability of the set of indicators. **Table 4** depicts the instrument's internal consistency to measure the latent constructs after removing the manifest variables: $x_2$, $x_3$, $y_2$, and $y_{13}$. Cronbach's Alpha results show an improved internal consistency of the indicators of the constructs [E] and [G] and no significant improvement for [M] after modification of the model. Besides, we see a slight reduction of the internal consistency of the overall sample, e.g., the decline of the global Cronbach's Alpha standardized from 0.915 to 0.905, even maintaining the reliability analysis on an excellent level. After those adjustments, we can understand and support the plausibility of the structures underlying the manifest variables and their relationship with the six latent variables of the theoretical model derived from that theory.

The SEM model in **Fig. 5**, reflecting the **Startup scenario ($\varphi_4$)** of the theory, was modified based on model adjustments and generated a refined model portrayed in FIG. 6.

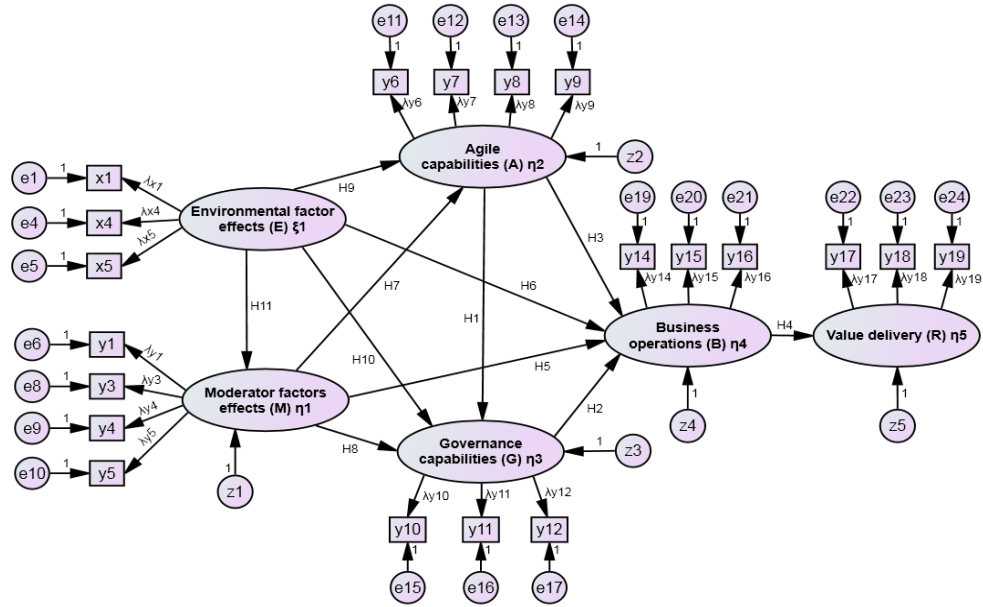

FIG. 6. SEM THEORETICAL MODEL BASED ON THE STARTUP SCENARIO ($\varphi_4$): ADJUSTED MODEL.

Such adjustments have only impacted some manifest variables (e.g., the Empirical Indicators), which are observational elements of the participant's perception of the phenomena under study. Those adjustments in the model did not jeopardize the plausibility assessment of the AGT's key constructs and core assumptions since the changes did not modify its core characteristics. The adjusted SEM models for every theoretical scenario designed for AGT are depicted in APPENDIX E (see part A of the figures).

### 4.3.3. Confirmatory Factor Analysis (CFA)

The objective of this modeling stage is to verify, using CFA, whether the correlational structure of this measurement sub-model reproduces the empirical evidence of the data sample. The CFA was performed on six measurement sub-models designated by the names of the constructs that sought to measure. We evaluate every sub-model using IBM® SPSS® Amos 20.0.

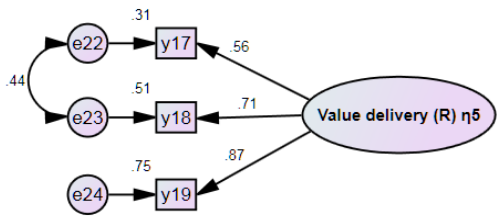

FIG. 7. ADJUSTED MEASUREMENT SUB-MODEL: VALUE DELIVERY [R].

During the CFA analysis, only two measurement sub-models related to the constructs [G] and [R] demanded some adjustments from the six latent constructs. For instance, in the first attempt, the measurement sub-model of *Value delivery* [R] did not present a good adjustment. Then, following the suggestions from Marôco (2014), we adopt the strategy to analyze the "*Modification Indices*" based on the CFA results from the data processing. Consequently, we correlated the measurement errors from the variables $y_{17}$ and $y_{18}$. We analyze each standardized correlation coefficient in keeping with Kline (2011). CFA results for this sub-model are depicted in **Fig. 7**, which corresponds to the screen of the fitted model with the display of standardized estimates. In fact, after model adjustment, all manifest variables have presented high factor weights ($\lambda \geq 0.5$) and appropriate individual reliability described by the squared multiple correlations ($R^2 \geq 0.25$). The measure sub-model demonstrated good quality fit: *Standardized Chi-square* ($\chi^2/\mathrm{df}$) = 0.001; *Comparative Fit Index* (CFI) and *Goodness of Fit Index* (GFI), both = 1.000; *Tucker-Lewis Index* (TLI) = 1.025; *Root Mean Square Errors of Approximation* (RMESEA) = 0.

However, the standardized correlation coefficient between the errors from $y_{17}$ and $y_{18}$ was 0.440. It can be interpreted as the existence of a factor (another manifest variable to measure the value of [R]) that is not included in the model, affecting these two variables in an equally strong manner in the same direction. From a theoretical point of view, "greater **utility** ($y_{17}$) embedded in products or services would require greater **warranty** ($y_{18}$) for the same ones" (Luna, 2015).

This adjustment strategy makes sense when we consider the existence of theoretical references about the correlation of these concepts. For instance, according to ITIL v3 (OGC, 2007), from the customer's perspective, value consists of two primary elements: *utility* or fitness for purpose and *warranty* or fitness for use. The customer perceives the **utility** of the service attributes that positively affect the performance of tasks associated with desired outcomes. In complement, the **warranty** is derived from the positive effect of being available when needed, with enough capacity or magnitude, and dependable in terms of continuity and security. *The utility* is what the customer gets, and the *warranty* is how it is delivered. Customers cannot benefit from something suitable for purpose but not for use, and vice versa. It is useful to separate the logic of *utility* from the logic of *warranty* for design, development, and improvement. However, they are complementary concepts strongly related to value delivery.

Considering all the separate controllable inputs allows for a broader range of solutions to the problem of creating, maintaining, and increasing value. Hence, we must scrutinize this relationship deeply, as well as other factors that might explain [R] in further studies. As illustrated, all the adjustments performed were adequately supported by theory, related works, and evidence gathered during the development of this research. After those adjustments, the measurement sub-models for every construct could reproduce the empirical evidence of the data sample appropriately.

### 4.3.4. Validity related to the constructs

Unfortunately, there is no single definitive test of the instrument's validity. Instead, usually investigated by SEM analysis, the so-called "validity related to the constructs" is separated into three components: (1) factor validity, (2) convergent validity, and (3) discriminant validity.

The (1) *factor validity* analyzes whether the indicators adequately measure the construct. In this aim, the standardized factor weights ($\lambda_{ij}$) are quantified, and it is assumed that the construct has factor validity when $\lambda_{ij} \geq 0.5$ for every manifest variable which describes it (Lomax and Schumacker, 2012; Marôco, 2014). APPENDIX H depicts the results of the factor validity of the constructs from this study. We also have verified the: (2) *convergent validity*, i.e., whether their correlations are, at least, moderate in magnitude: there has been convergent validity for every construct; *composite reliability* (CR), which evaluates the degree to which the empirical indicators from a constructed measure the latent concept related to it: all constructs from the model have very good CR values; and (3) *discriminant validity*, to identify which sets of indicators (factors), are substantially different for each construct: the results confirm the discriminant validity for every construct.

### 4.3.5. Models estimation for validity analysis

After checking the measuring instrument reliability and validity, we have estimated the complete model and the support or rejection of the causal hypotheses to the phenomena under study in every theoretical scenario from the AGT.

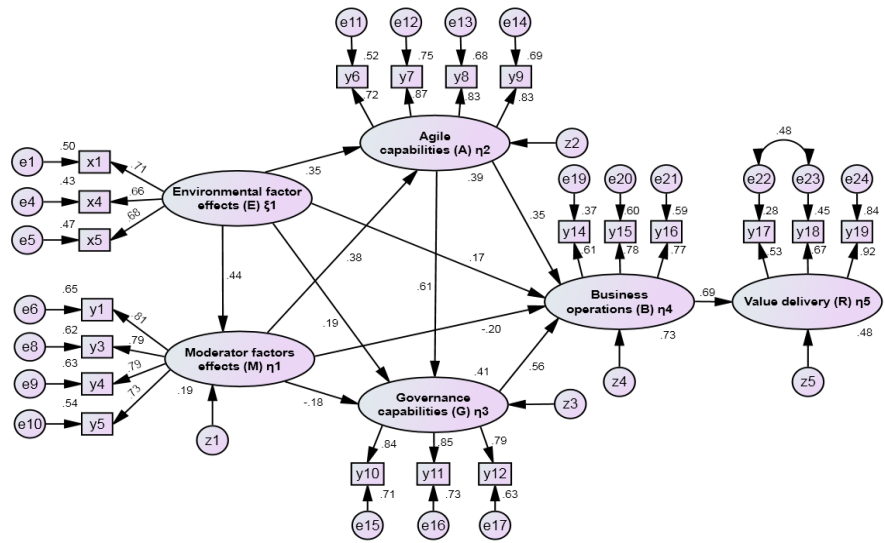

**FIG. 8.** ADJUSTED SEM THEORETICAL MODEL BASED ON STARTUP SCENARIO (φ4).

After making the CFA of measurement sub-models, following the methodology "two-step", we have continued evaluating the complete model for every AGT theoretical scenario using the procedure depicted in Section 3.1.2. This step considers the combined treatment of the six measurement sub-models and the structural sub-model as the second stage of SEM modeling, allowing assess the causal hypotheses of the phenomena under study.

We checked the adjustment quality of the measuring sub-models by the respective indices at CFA analysis. At this stage, we have attempted to assess the plausibility of the complete models. Therefore we used the same indexes applied to the measurement model, incorporating parsimony indexes. These indices check whether the best-adjusted model (by adding parameters or relationships) is also *parsimonious*[16] (simple model).

In each theoretical scenario, we assess the hypotheses preconized by the AGT by employing the *Path Analysis* approach allowing us to test altogether from $H_1$ to $H_{16}$ (Marôco, 2014). In the case of the hypothesis $H_{16}$, we have assessed the supposed mediation between the constructs [A] (predictor variable) and [B] (dependent variable) through [G] (mediator variable), adopting two statistical tests: (1) Sobel Test for Mediation (Marôco, 2014; Sobel, 1982); and, (2) Bootstrap Test for Mediation (Marôco, 2014) (p. 155).

*4.4. Path Analysis*

This section discusses the **Startup Scenario ($\varphi_4$)** estimation as an exemplification of the procedures adopted for every model derived from the theoretical scenarios (see Section 3.1.2). The model ($\varphi_4$) estimation results are depicted in **Fig. 8**, and its adjustment indices are presented in **Table 5.** For analyzing models' estimations results, we are using the SEM quality indices for the model fit described by Marôco (2014) (p. 55) and Mulaik (2009), depicted in **Table 3**.

TABLE 5. **QUALITY ADJUSTMENT INDICES FROM THE MODEL BASED ON STARTUP SCENARIO ($\varphi_4$), (N=118).**

| Group Analysis | Indices | Obtained values | Reference values for Analysis[17] | Analysis |
|---|---|---|---|---|
| *Fit tests* | *Chi-square ($\chi^2$)* | 202.300 | lower is better | - |
| | *Degrees of freedom (df)* | 158 | $\geq 1$ | OK |
| | *p-value* | 0.010 | >0.05 | OK |
| *Absolute Indices* | *Standardized Chi-square ($\chi^2/df$)* | 1.280 | <3 | OK |
| | *Root Mean Square Error of Approximation (RMSEA)* | 0.049 | <0.10 | Good fit |
| | *Goodness of Fit Index (GFI)* | 0.857 | >0.90 | Acceptable fit |
| *Relative Indices* | *Comparative Fit Index (CFI)* | 0.962 | >0.90 | Good fit |
| | *Normed Fit Index (NFI)* | 0.851 | >0.90 | Acceptable fit |
| | *Tucker-Lewis Index (TLI)* | 0.954 | >0.90 | Good fit |
| *Parsimony Indices* | *Parsimony GFI (PGFI)* | 0.644 | >0.60 | Good fit |
| | *Parsimony CFI (PCFI)* | 0.800 | >0.60 | Good fit |
| | *Parsimony NFI (PNFI)* | 0.708 | >0.60 | Good fit |
| | | Solution is admissible | - | Acceptable Fit |

Aiming to improve the model quality adjustment indices, we have correlated the measurement errors from variables $y_{17}$ and $y_{18}$ based on a discussion carried out in Section 4.3.3. Observing the data in **Table 5,** we realize that, except for the GFI and NFI, all other indices assessed to meet the references suggested in the literature (Hair et al., 2009; Kline, 2011; Lomax and Schumacker, 2012; Marôco, 2014). Even so, although the GFI and NFI values were less than 0.9, they were appreciably high: above 0.8.

In this study, in keeping with Zikmund (2003) and Bollen (1989), we can imply that the GFI and NFI values lower than 0.90 are a consequence of a sample size that is not so large, but not necessarily, due to a bad model fit. According to Marôco (2014) and Marsh, Balla, & Hau (1996), CFI and TLI are more appropriate for assessing model quality adjustment indices in a small sample size. Indeed, both values calculated to CFI and TLI have a "good fit", as depicted in **Table 5**.

We can conclude that the model adequately reproduces the correlation of empirical data structure, allowing us to classify the overall analysis of the quality adjustment indices as acceptable for the purpose it is intended for this empirical study: to assess the plausibility of the AGT **Startup Scenario ($\varphi_4$)**.

According to Marôco (2014), we must analyze the value of the path coefficients (β) to test each hypothesis from the current model, i.e., the "*standardized regression weights*" for related constructs. Likewise, we consider their estimated probability of getting a sample value this far from zero if the population value is zero (p), i.e., the "level of significance for each regression weight" (Efron and Tibshirani, 1994).

When the estimated probability (p) is statistically significant for each hypothesis, its path coefficient (β) is recognized as a valid causal linkage between the constructs under analysis. In other words, we reject β for p > 0.05. Regarding the hypotheses testing, in the scenario under analysis, there are 12 hypotheses from 16 hypotheses depicted from the emerging theory, which we intend to test. The test result and analysis for each hypothesis from the **Startup Scenario (φ4)** are depicted in APPENDIX I.

---

[16] They are models that seek to explain data with a minimum number of parameters, or predictor variables, having great explanatory predictive power.
[17] According to Marôco (2014) and Mulaik (2009).

Finally, the standardized regression coefficients (β) of the structural models, and the significance levels (p) associated with each hypothesis in every scenario analyzed, were assessed using the same approach employed to test **Startup Scenario (φ4)**. APPENDIX J depicts the goodness fit indices of the SEM model for each scenario. The quality adjustment indices for every model derived from theoretical scenarios designed for AGT are depicted in APPENDIX J. The results of the estimation for every model derived from theoretical scenarios designed for AGT are depicted in APPENDIX E (see part B of the figures), as well as its adjustment indices are presented in APPENDIX K.

Also, the significance of the direct effects was obtained by Bootstrap simulation. In this research, following the customary practice of SEM studies, the *Critical Ratio* (CR) is treated as a "*t value*"[18], which is associated with the "*t-test*" used to check if the factor weights between two variables are significantly different from zero at a given probability level. Typically, according to Marôco (2014), regarding CR, they have accepted values of $t > 1.96$, i.e., $p-value < 0.05$ (two-tailed), which means that the findings have significance at a 97.5% confidence level, in keeping with Vincent & Weir (1994). The regression coefficients of the models are depicted in their standardized form.

Aiming to illustrate the rationale of the hypothesis test, adopting the SEM Path Analysis approach, we detail the analysis developed for the hypothesis $H_1$ in φ4, as follows:

$H_1$ <u>can be supported</u> ($\beta_{H_1} = 0.606$, $p_{H_1} < 0.001 ***$) *because the probability of getting a critical ratio as large as 4.379 in absolute value is less than 0.001. In other words, the regression weight for [A] in the prediction of [G] is significantly different from zero at the 0.001 level (two-tailed). Further, when [A] goes up by one standard deviation, [G] goes up by 0.606 standard deviations. Hence, analyzing the current scenario, based on the data sample gathered by this empirical study, we can substantiate that "the agile capabilities [A] have a positive influence on governance capabilities [G]" in the Startup Scenario (φ4) with enough statistical significance.*

However, in practicality, we will not detail the analysis for every hypothesis in each theoretical scenario. We will only depict the test results for each hypothesis. Likewise, we provide the Quality Adjustment Indices and the Hypotheses Test for every theoretical scenario, respectively, in APPENDIX J and APPENDIX K. We also discuss the overall results of this procedure in Section 5.

## 5. DISCUSSION

In this section, we discuss the study findings. The model fit varies from scenario to scenario, as seen from the statistical results of our empirical data analysis. The following sections discuss how these results address our research question.

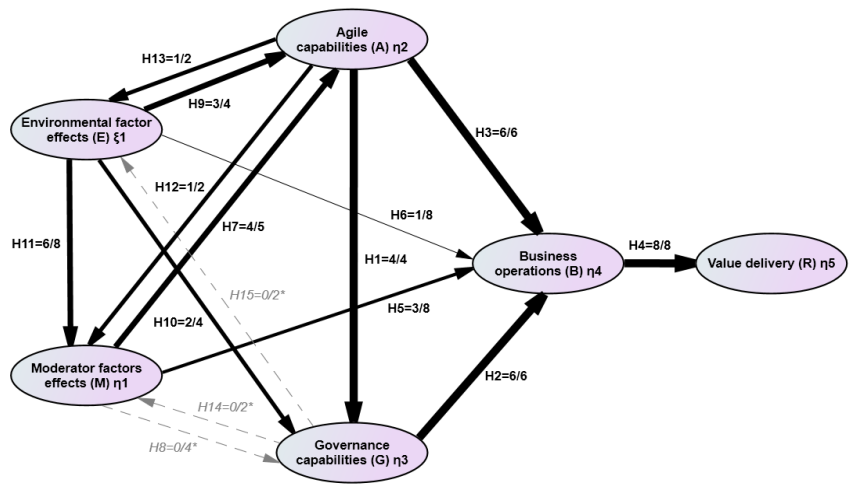

**FIG. 9.** OVERALL HYPOTHESIS ANALYSIS FROM THE STRUCTURAL EQUATIONS MODEL BASED ON DYNAMIC SCENARIO (φ$_n$), (N = 118).

### 5.1. Data Synthesis: analysis of the study hypotheses

**Fig. 9** depicts graphically the hypotheses represented by the unidirectional arrows (relationships between constructs). Each arrow displays the hypothesis code and the number of times the hypothesis was supported, separated by the "slash symbol" ("/") from the number of times the same hypothesis was tested in all scenarios. For instance, $H_1$ was supported in four scenarios out of the four scenarios in which it was tested. The hypotheses supported are indicated by solid black lines, where the line width is associated with the frequency with which each hypothesis was supported in the various scenarios studied. In turn, the hypotheses not supported (which have no meaningful relationships) are represented by gray dashed lines.

---

[18] Adopting the notation suggested by Marôco (Marôco, 2014), during hypothesis testing, we have used the notation |Z| for characterizing "*t value*", as depicted in the Sobel test for $H_{16}$.

### 5.1.1. Hypotheses with strong support

Our statistical modeling provided evidence in support of the **hypotheses $H_1$, $H_2$, $H_3$, $H_4$, and $H_{16}$ in all eight scenarios.** These hypotheses are related to the key constructs ([A], [G], [B], [R]) and core assumptions of the theory. The results lead to the following:

(1) To support the improved coherence of the agile governance concept published in (Luna et al., 2016), which characterize agile governance as "*the capability of an organizational context[19] to sense, adapt and respond to changes [A] in its environment, in a coordinated and sustainable way [G], faster than the rate of these changes*". This definition considers the relation among the constructs [A] and [G] with each other and with the business operations [B], aiming to generate the greatest possible value delivery. In other words, this concept is closely related to **all those supported hypotheses**.

(2) To support the characterization and differentiation between the *agile governance approach* and *specific agile approach* published in (Luna et al., 2014) and briefly discussed in Section 2. The latter has its influence limited to a localized result, usually a few stages of the organization's value chain (Porter, 1985). While agile governance emerges as the balanced application of agility upon the system responsible for sensing, responding, and coordinating the entire organizational body: the governance (or steering) system. The *specific agile approach* is characterized by $H_3$, while $H_1$, $H_2$, and $H_{16}$ depict the *agile governance approach*.

(3) To support the plausibility of the AGT and its core components, leading to improved verisimilitude in aspects such as:

   a. **1st Law:** as depicted in APPENDIX D, is directly related to $H_1$, $H_2$, $H_4$, and $H_{16}$, which states that "*agile governance is the result of the coordinated combination of agile capabilities [A] with governance capabilities [G], in which the blend of these two capabilities increase the level of business operations [B] (in quality and/or productivity), which in turn increases the value delivered [R] (to the business)*" (Luna et al., 2020). This phenomenon was found in not recent and new literature findings such as:

   i. Aiming to establish a resilient infrastructure for mission-critical services, Johns Hopkins University Applied Physics Laboratory researchers propose cybersecurity mechanisms with autonomic properties that allow them to become more self-aware and react in near real-time to attacks and failures, considering the scope of multi-UAS air-traffic control (Maurio et al., 2021).

   ii. Public administration decision-makers are seeking ways to implement agile governance mechanisms in the context of smart cities to predict and reduce the vulnerability of these cities. Also aiming to help "citizens to play a leading role in updating the situation's current trends" in towns and responses related to the post-COVID-19 scenarios, allowing them "actively interact with public administration decision-makers" (Founoun et al., 2022).

   iii. The Policymakers and policy executors of the Dutch Tax and Customs Administration (DTCA) were searching for ways to achieve higher levels of flexibility and agility in their business process management systems, performing "a flexible and agile policy implementation" (Gong and Janssen, 2012).

   iv. Still, in the context of policy-making, people are perceived as "the driving force behind policy" by Black's (2022) study. Agile governance is described as an instrument to overcome "transport and mobility challenges in Japan". The Japan Ministry of Economy, Trade, and Industry (2021) adopts an AG approach to "carry out ongoing analysis of the social situations the stakeholders (including governments, businesses, individuals, and communities) find themselves in, define the goals they seek to achieve, design the various systems for achieving these goals and carry out ongoing dialogue-based assessments of outcomes to make improvements to these systems".

   v. In enterprise social media, Baptista et al. (2017) pointed out reflexiveness as a new capability for an open strategy to move organizations toward giving employees a higher degree of independence and ownership of their work and creating feedback loop processes to shift organizational governance systematically to a responsive approach.

   vi. Practitioners and scholars have been developing ITG "mechanisms to improve sensing, deciding, and responding capabilities in turbulent business environments" for Digital Governance seeking to produce resilient competitive organizational advantage and allow "agile responses in dynamic competitive environments" (Vaia et al., 2022).

   vii. The U.S. Government has been looking "accelerate the policy development process for digital government" in four U.S. federal agencies (Parcell and Holden, 2013);

---

[19] Considering a team, a project, a business unity (sector of a company), an entire enterprise, or even a multi-organizational environment, as already discussed in Section 2.

viii. Auditors have been reflecting on the "agile auditing" concept by adopting principles and values from the Agile Software Development Manifesto. They seek to "eliminate, or at least mitigate, the negative stigma which many auditors have grudgingly accepted", aiming to provide more value to the organization and its stakeholders due to auditing processes (Truong, 2020).

ix. A Dutch firm, together with an offshore site in India, which develops software for the finance market, was applying a "Multisite governance model for Global Software Development" to analyze their multisite governance activities adopted and adjusted based on the Scrum methodology (Noordeloos et al., 2012);

x. Logistics professionals and scholars have been developing "agile sustainability governance mechanisms" for emerging market supply chains on advanced economy multinational enterprises (AMNEs). Soundararajan et al. (2021) report initiatives based on a processual approach, seeking to implement flexible, agile, and adaptive instruments to produce more robust and innovative solutions to logistics involved in those scenarios.

xi. Israeli Air Force has "become more effective the governance of agile software teams for large-scale projects" for their defense industry (Talby and Dubinsky, 2009);

xii. Center for Urban Studies at the University of Amsterdam reports its experience developing "agile organizational capabilities". It is aimed at urban planning in Amsterdam's Bicycle Program through an "interactive work proposal", considering "collaboration, experimentation, continuous improvement, and organizational learning" for the public administration decision-makers and their stakeholders in transportation and urban mobile practices (Hahn and te Brömmelstroet, 2021).

xiii. Eight worldwide organizations of distinct size, industries (e.g., Insurance, Bank, Automotive, Health, Government, Logistics, Production, and Retail), and experience with Enterprise Architecture Management (EAM) pointed out that EA Governance increases "strategic agility", bringing benefits, such as the "ability to deal with changes", and being "more responsive" (Ahlemann et al., 2020);

xiv. Educators have proposed an "agile-based instructional method" grounded on "agile processes that emphasize self-direction, collaboration, and lightweight procedures" for "fostering innovation in university teaching and learning", tested at Universität München with more than 150 individual innovation and online learning projects (Wirsing and Frey, 2021).

xv. To deal with the COVID-19 outbreak with limited information and confronting many uncertainties, considering the "ability of countries around the world to respond in an agile and adaptive manner", particularly regarding the "timing of policy measures", the "level of decision centralization", the "autonomy of decisions" and the "balance between change and stability" (Janssen and Voort, 2020), among others examples.

b. **2$^{nd}$ Law:** as depicted in APPENDIX D, is intimately related to **H$_3$** and asserts that "*a specific agile approach arises when agile capabilities [A] are applied directly on business operations [B] (without the mediation of governance capabilities [G]), activating or intensifying an increase in [B], which in turn increases the value delivery [R]*" (Luna et al., 2020). Indeed, this phenomenon was observed abundantly in literature. It can occur in the form of several approaches, such as agile software development (Henríquez and Moreno, 2021; Wang et al., 2011), agile project and portfolio management (Ploder et al., 2022; Thomas and Baker, 2008), agile manufacturing (Khalfallah and Lakhal, 2021; Sun et al., 2005), and agile enterprise and agile enterprise architecture management (Alzoubi and Gill, 2022; Bider et al., 2013), among others. Besides, were also found *specific agile* applications in new contexts, such as public management (Balakrishnan et al., 2022), enterprise social media (Pitafi et al., 2020), logistics (Black, 2022), health care (Batayeh et al., 2018), and financing management (Tou et al., 2020), to cite a few.

c. **6$^{th}$ Law:** as depicted in APPENDIX D, is closely related to **H$_4$** and expounds that "*influence on business operations [B] will generate directly proportional effects on value delivery [R]*" (Luna et al., 2020). This sixth law receives as input the resultant effect of every law of AGT and their influence upon business operations [B] and relays their consequences on value delivery [R]. [R] conceptualizes the ability to generate results for the business through the delivery of value, which includes all forms of value that determine the health and well-being of the organization in the long run. Further, these results must be expressed by delivering products and services, which are outcomes of business operations [B]. Moreover, these results can be mainly perceived by the persistence of the benefits arising from those products and services. Might be examples of these benefits: customer satisfaction, business efficacy, humanitarian aid, and the welfare of citizens, considering a broad spectrum of the audience that depends on every core business, such as shareholders, customers, employees, partners, suppliers, and society.

d. **Theory's premises published in** (Luna et al., 2015):

    i. **Premise 1:** introducing agile governance as the balanced application of agility upon the governance system, related to the already mentioned differentiation between the *agile governance approach* and a *specific agile*

> *approach* published originally in (Luna et al., 2014), the prior one aligned with the 1st Law of AGT and characterized by $H_1$, $H_2$, and $H_{16}$, and the latter lined up with the 2nd Law and depicted by $H_3$.
>
> ii. **Premise 2:** Concerning the positioning of the phenomena, we imply the *agile governance as a socio-technical phenomenon* positioned in a *chaordic* range between the innovation and emergent practices from agile (and lean) philosophy and the *status quo* of the best practices employed and demanded by the governance issues, as already discussed in Sections 2 and 4.2. The socio-technical nature of agile governance is substantiated due we are handling the understanding of the intersections between technical and social aspects: considering people as agents of change in organizations in contexts where technology is a key element (Luna et al., 2014). This assumption was also supported by the study results and is discussed in Section 4.2.
>
> iii. **Premise 3**: Finally, the third premise is the definition of agile governance as a broad concept closely related to **all those supported hypotheses**, as already discussed on the topic (1), and its meta-principles and meta-values proposed in (Luna et al., 2014) and (Luna et al., 2016).

(4) Based on the results of mediation analysis from $H_{16}$, we can also infer that the effect of Agility [A] on Business Operations [B] has a higher intensity when applied through the mediation of Governance capabilities [G] than when it is applied directly to Business Operations [B]. This result also supports **Premise 1** in (Luna et al., 2015).

### 5.1.2. Hypotheses with limited support

Other **hypotheses were supported in some scenarios, such as $H_5$, $H_6$, $H_7$, $H_9$, $H_{10}$, $H_{11}$, $H_{12}$, and $H_{13}$**. These hypotheses are related to the interaction among the key constructs and the surrounding (disturbing and restraining) constructs: [E] and [M], respectively. Therefore, we could infer that the supposed "instability" regarding some hypotheses being supported in some scenarios (and others not) might be related to the choice of the set of empirical indicators to measure those constructs. It was discussed in Section 2 and APPENDIX B, considering that the more generic those constructs were, the more inaccurate our measure of them because they are those theory units that are more sensitive to the nature of the organizational context boundaries (Luna et al., 2015). This inference is supported by the explanatory power identified in each scenario for those constructs: they often are the constructs with lesser *squared multiple correlations*. For instance, in Dynamic Scenario ($\varphi_n$), it is estimated that the predictors of [E] explain 14 percent of its variance. In other words, the error variance of [E] is approximately 86 percent of the variance of [E] itself.

On the other hand, it would be fair also to infer, without significant elaborations and simply interpreting these results as correct, that the theory does not hold concerning each hypothesis in the respective theoretical scenarios where they were not supported. Regardless, we can imply that the hypotheses from this second group are supported in the respective theoretical scenarios where they were supported, pointing out the behavior depicted by these hypotheses as plausible in those scenarios. According to APPENDIX D, these results influence the verification of the plausibility of the 3rd ($H_5$ and $H_7$), 4th ($H_6$, $H_9$, $H_{10}$, and $H_{11}$), and 5th ($H_{12}$ and $H_{13}$) Laws of AGT, namely:

- **Law 3** declares, "*there are internal moderator factors whose effects [M] can inhibit or restrain the agile capabilities [A] and governance capabilities [G], or even reduce business operations [B], which in turn decreases value delivery [R]*".

- **Law 4** describes that "*there are environmental factors whose effects [E] can disturb the organizational context. [E] influences the effects of moderator factors [M], agile capabilities [A], governance capabilities [G], and business operations [B], which in turn, on some level affect value delivery [R]*".

- **Law 5** articulates that "*the combined and coordinated coupling of agile capabilities [A] and governance capabilities [G] reduces the effects of environmental factors [E] and moderator factors [M] on the organizational context. Combined [A] and [G] contribute to decreasing the inhibition, restriction, or disturbance in the organizational context, and decreasing their harmful effects on business operations [B] over time, which in turn increases value delivery [R]*".

However, further studies are necessary to develop a more in-depth analysis and reach more consistent conclusions.

### 5.1.3. Unsupported Hypotheses

Finally, only the **hypotheses $H_8$, $H_{14}$, and $H_{15}$ were not supported in any scenario**. None of these hypotheses had enough statistical significance, according to the data collected in this study. In other words, no evidence was found to confirm the negative influences of the Moderator factors effects [M] over Governance Capabilities [G] ($H_8$). Neither was identified any evidence that supported the positive influence of Governance capabilities [G] over both: Moderator factors effects [M] ($H_{14}$) and Environmental factors effects [E] ($H_{15}$).

In the case of $H_8$ and $H_{14}$, it seems to make sense. It opens a perspective for reflection on how to improve the description of the behavior of the phenomena involved that Governance capabilities [G] may not drive factors beyond the organization's internal boundary (i.e., the theory's open boundary that determines the organizational context). However, it was unexpected to note that Governance capabilities [G] cannot drive the Moderator factors effects [M] ($H_{15}$) since they reside within the boundary of the organizational environment.

Adding to the inference discussed in the previous paragraph, we could imply other reasons for these results, such as (1) this result might be a consequence of the multiple organizational contexts (see Section 4.2) shuffled into the data sample, leading us to infer that the creation of a data set for each theoretical scenario may be a strategy to remove this potential influence in future studies; and/or, (2) this evidence may still be an effect from the question wording (quite generic questions for some empirical indicators) in the questionnaire because it was developed based on the broader scenario for theory instantiation, the **Dynamic Scenario ($\varphi_n$)**, as discussed in Section 3.1; and/or even, (3) a misinterpretation caused by an inaccurate characterization of some empirical indicator, allowing the respondent to answer some question without precisely understanding the context related to what it was intended to measure. Seeking to be impartial, another way to evaluate these results is that the interpretation is correct and that the theory does not hold concerning these three hypotheses. In this case, in practice, as depicted in APPENDIX D, these results would not allow verification of the plausibility of the 3$^{rd}$ (**$H_8$**) and 5$^{th}$ (**$H_{14}$ and $H_{15}$**) Laws of AGT. Regardless of any conjecture, further studies are needed to verify these three hypotheses and the scenarios where they take place.

At the same time, lessons learned by this study about the complexity of the phenomena under investigation and the specificity to be considered for each subsample suggest improvements to methodological procedures and the data collection instrument. For instance, in future studies, the questionnaire shall be tailored according to: (1) each *organizational context* because context boundaries delimit the type of empirical indicator that shall be chosen to measure constructs (especially regarding surrounding constructs such as [E] and [M]); (2) adequacy of the investigated *theoretical scenario* to the respondent actuality, i.e., displaying specific examples and question-wording according to the chosen scenario by them; and, (3) using the positioning of the distinct organizational contexts on the *chaos and order scale* (see Section 4.2) to both shape the questionnaire, and to analyze properties from the theory, such as generalization, causality, explanation, and prediction. The results obtained by this study are substantially relevant for advancing the understanding of if (or how) AGT works in practice and improving the knowledge of how to develop more reliable instruments to evaluate it in future studies.

## 5.2. Results interpretation

Although not all hypotheses have been supported in every scenario, the Confirmatory Factor Analysis (CFA) and Structural Equation Modeling (SEM) analysis point out theory verisimilitude, supporting the hypotheses related to the key constructs and core assumptions of the AGT in every theoretical scenario.

Following the study described in this manuscript, we can now positively answer the *research question* stated in Section 1: *"Are the key hypotheses and core assumptions of the Agile Governance Theory supported by practitioners' experience?"* The evidence brought by this study supports the recognition of the plausibility of the theory components and central hypotheses.

However, considering the *strength of evidence* and *limitations* experienced by this study, which will be discussed in the subsequent sections, the results also indicate that further studies are necessary to reach a trustworthy theory to describe and analyze the agile governance phenomena. Therefore, these findings lead us to infer that this empirical study has successfully assessed the theory's plausibility. Still, much work must be done to evaluate (and, eventually, enhance) the theory. This study also brought many lessons that researchers can use to refine and expand future studies on this topic.

## 5.3. Strength of evidence

We adopted the GRADE working group definitions to assess the strength of our evidence based on our research method (Atkins et al., 2004). As stated in GRADE, the strength of evidence can be defined through the consolidation of four keystones: (1) **study design**, (2) **study quality**, (3) **consistency**, and (4) **directness**. These key elements are assessed in four levels of grading: (a) *High*: Further research is quite improbable to change the reliability in the estimation of effect; (b) *Moderate*: Additional research is probably to have a relevant impact on the confidence in the estimation of effect and might modify the estimation; (c) *Low*: Further research is quite probable to have a relevant impact on the reliability in the estimation of effect and is probably to modify the estimation; and, (d) *Very low*: Any estimation of effect is very unreliable. The GRADE system considers evidence established on study design, procedures followed during study execution, and the data collected, analyzed, and synthesized. Nevertheless, by evaluating the other key elements mentioned, the initial assessment could be revised and amended according to the identification of inconsistencies or high-quality observational inquiries, for instance.

Concerning the **study design** reported in this report, it was framed by a *research framework* developed to guide this study (depicted in **Fig. 2**) for quantitative research. It includes advanced statistical methods, such as Structural Equation Modeling (SEM) and Confirmatory Factor Analysis (CFA), to conduct a cross-sectional research Explanatory Survey (Groves et al., 2013) and assess the theory. Despite the efforts, we believe we can get more accurate results for future studies by conducting controlled experiments and refining our data collection instruments. The initial classification of the overall evidence in this review grounded on study design is **moderate**. Now, we will address the other key elements of the studies in the evidence base.

Regarding the **study quality**, composed of a *cross-sectional explanatory survey* for quantitative analysis, we argue that the study was thoroughly planned and has its methods, audience, and approach described and registered by a study protocol. The participants and subject groups were selected from a predefined profile (see Sections 3.1 and 4) described explicitly and coherently with the maturity of the phenomena under study and considering clear and objective criteria, i.e., the AGT's external boundary-determining criteria. However, we recognize that the study participants were not selected randomly, which could give better reliability to their results. It happened because, considering agile governance as a *developing* field of study, we have had a small sample available for work. Based on these findings, we imply that there are restraints to the studies' quality that unavoidably increase the risk of bias or misconception. Hence, we must be cautious about the studies' confidence.

Concerning the **study consistency**[20], we not found significant differences related to the alignment among its findings, meaning good correspondence of effect estimates across data, considering the diversity and complementarity of the data sample from the *cross-sectional explanatory survey* described and discussed in Sections 4.1, 4.2, 4.3, and 4.4.

Regarding **directness**[21], the results presented in Section 4 and discussed in Section 5 about the *cross-sectional explanatory survey*, we can advocate that they point to theory verisimilitude, supporting the hypotheses are related to the key constructs and core assumptions of the theory in every theoretical scenario. Although further studies are needed to clarify some points of the theory that the results of this study could not support.

Analyzing these four components of the study, we identified that the strength of the evidence concerning this research is **moderate**. Therefore, it is probable that further research in this area may impact and modify the reliability of the theory test results. Despite these limitations, the research presented sheds some light on how to broaden our understanding of the agile governance phenomena.

### 5.4. Limitations

To scrutinize the limitations of this research, we discuss below: construct validity, internal validity, and external validity.

#### 5.4.1. Construct validity

Usually, the major research limitations reside in the methodological approach adopted to carry out the research. In this aspect, this work adopts a quantitative study to assess a theory for analysis and description (Gregor, 2006) that emerged from mixed methods that combine qualitative and quantitative research approaches. The risk related to the methodological approach has been reduced by the development of the *research framework* depicted in **Fig. 2**, the procedures characterized in **Table 1**, and the research design discussed in Section 3. Furthermore, the *research framework* was underpinned by a specific protocol, followed in a disciplined manner, generating results reported in this report.

Despite this attempt, we believe that additional improvements in the data collection instrument can enhance data quality to be gathered in further studies, e.g., based on the commonalities analysis (Section 4.3.2), we had to remove some manifest variables from the models because their common factor did not adequately explain the variance of these items (e.g., $x_2$ and $x_3$, their commonalities were ≤ 0.50).

Another point related to the SEM analysis regarding the quality indices adjustment of the models is the *size sample*. Some indices behave erratically across estimation methods under a small sample size. For instance, the Normed Fit Index (NFI) is not a good indicator for evaluating model fit when a small sample size. Aiming to minimize this potential bias, we have applied the Bootstrap procedure with 2000 resampling.

We recognize that the list of empirical indicators adopted for this study is not a definitive list that should be used to measure the constructs of the theory under either circumstance. The set of empirical indicators must be adjusted and refined for each concrete case (theory instantiation) according to the *organizational context* under study. One of the lessons learned from this study is that the accuracy of the hypotheses test depends on the suitability of the choice of empirical indicators (see Sections 2, 4.3.2, and 5.1). In this study, we chose a set of indicators intended to represent most situations observed during the research development, which we call *a general view of the theory*.

We knew many other indicators could be suitable for measuring the theory's constructs. However, after this study, we become even more conscious of the relevance of choosing empirical indicators according to specific variations of (1) the nature of the organizational context under analysis, (2) the influence of the surroundings experienced by this context, and (3) the awareness on agile governance identified in this context, according to the characterization performed by the system states described in AGT.

#### 5.4.2. Internal validity

Regarding the methods adopted in this research: our sampling composition has adopted a mix of purposive sampling types (Patton, 1990), which is usually considered less rigorously than a random sample (see Section 4.1). However, our choice can be justified by the nature of the phenomena under study as a *developing topic*, considering the publication of the first definition of agile governance by Asif Qumer Gill (Qumer, 2007). We had difficulty applying random sampling in this peculiar context, where the representative agents of these phenomena were rare and difficult to identify. So, they needed to be chosen intentionally. Considering the constraints faced by the research during the empirical study, the survey data sampling strategy is transparent. However, future studies should be more selective to advance in theory assessment. They must examine data sampling based on organizational contexts recognized by agile governance applications, such as in public administration, business administration, and industry, such as agile software development projects. We can infer that real empirical evidence can be found more abundantly in those contexts than among scholars or authors of agile literature.

---

[20] According to Atkins et al. (Atkins et al., 2004) it refers to the similarity of estimates of effect across studies and or data samples.

[21] According to Dybå & Dingsøyr (Dybå and Dingsøyr, 2008) it concerns the extent to which the people, interventions, and outcome measures are similar to those of interest.

Besides, despite all the effort and energy employed in the empirical study data collection, we have a small survey sample size (N=118). Despite that, the survey sample size has enough representativeness and statistical significance to allow us to develop statistical analysis. Regardless, we infer that due to the methodological diligence (rigor) followed by this research, the obtained results are at least a representative sampling of the phenomenon under study, as discussed in Sections 3.1.4, 4.1, and 4.2.

We can also infer as a potential bias the influence of some "explanatory variables" that could have affected the adjustment of quality indices from the models derived from each theoretical scenario depicted from theory can be the reason, namely: (1) the participant's experience related to agile governance phenomena; (2) the organizational context chosen to answer the questionnaire; (3) the self-evaluating positioning into Chaos & Order Scale; and, (4) the choice of the best suited theoretical scenario. Aiming to test that supposition in future studies, using a larger sample, we should analyze those subsamples (groups of cases) separately and compare the results. Another alternative to reduce that potential bias could develop specific questionnaires, adopting a distinct group of empirical indicators (contextualized) for every combination of those "explanatory variables" in future studies.

Moreover, the data collection instrument (questionnaire[22]) has been evaluated, with more than twelve[23] representative agents of the phenomena under study, through a Survey Pilot (see Section 3.1). Despite that, based on the results of this study, we understand that we must refine the wording of the questions and develop some concepts more clearly, in future studies. This effort to validate and refine the questionnaire tries to reduce the study's threat that if the participants do not understand the question, we cannot be guaranteed that they answered the survey correctly.

### 5.4.3. External validity

Concerning the generalizability of the findings, we have some limitations related to the survey sample comprising its significant amount (87.3%) of "typical cases" (see discussion in Section 4.1). Therefore, with the predominance of the typical case sampling, we cannot use the sample to generalize a population. However, the sample could be illustrative of other similar samples, according to Patton (1990).

Finally, none of the discussed limitations or potential biases related to this section substantially threaten the results of this research. We outline the study limitations to highlight possible improvements in future research that assesses and refines the theory and to share with other researchers the learning lessons from our study, aiming to help them improve their theory assessment studies.

## 6. CONCLUSION

This section summarizes our key contributions in context with our aims and research question. We conclude with an agenda for further research.

Considering that the core theory's hypothesis ($H_1$, $H_2$, $H_3$, $H_4$, and $H_{16}$) can be supported by an empirical study that reasonably implies the AGT as a plausible theory. These five hypotheses comprise a set of guidelines for governing organizations based on agile teamwork, which can be employed in several areas, such as software development, business management, or public administration. In turn, considering that the AGT is a plausible theory, we imply that the guidelines summarized in Section 5 can be important for developing a real impact in which agile governance can be understood and implemented by individuals, leaders, and managers throughout the entire organization.

### 6.1. Analysis of the research goals

This study aimed to assess the plausibility of the Agile Governance Theory as a consequence of the test of its *key constructs and core assumptions*. To this end, we conducted an empirical study in which 118 participants completed an explanatory survey.

The answers to the survey were analyzed using *Confirmatory Factor Analysis (CFA)*, evaluated the measurement sub-model, and verified the goodness of fit of the eight theoretical models related to the correlational structure of manifest[24] variables from the data sample collected during the study. Results of the CFA indicated that the correlational structure from each one of those eight theoretical scenarios in the respective measurement sub-models reproduces the empirical evidence of the data sample adequately. It can be supported when we consider that the *Overall Goodness of Fit* from every scenario was assessed as a "Good Fit" or "Acceptable fit" (see APPENDIX J).

Finally, we applied *Structural Equation Modeling (SEM)* to test the proposed hypotheses for each of the eight theoretical scenarios derived from the theory. The overall analysis of the eight scenarios is described in Sections 4.3, 4.4, and 5.1, as well as depicted in APPENDIX J and APPENDIX K. In short,

---

[22] Even by Dubin's criteria of *validity* and *instrument reliability*, we added to them the other two criteria, such as *clarity* (wording) and *relevance* for each question.
[23] We invited 15 experts, but only 12 of them completed the questionnaire (pilot) testing procedures in form and content.
[24] A manifest variable is a variable that can be directly measured or observed.

1) **five hypotheses were supported in all scenarios** ($H_1$, $H_2$, $H_3$, $H_4$, and $H_{16}$), for instance:
   a) The agile capabilities [A] have a positive influence on governance capabilities [G] ($H_1$). Denoting that [A] enhances the application of [G]. For instance, agility at the business level demands capabilities such as *flexibility*, *responsiveness*, and *adaptability*, which are examples of agile capabilities [A]. In turn, those [A] generate better results when applied by the combination of governance capabilities [G], such as *policymaking*, *decision-making*, *accountability*, *control*, and *responsibilities*: producing new and innovative approaches, such as *agile policymaking* (Howlett and Ramesh, 2022; Parcell and Holden, 2013), *agile decision making* (Borgman et al., 2013; Kokol et al., 2022), *lean public services* (Carter et al., 2011; Klein et al., 2022), among other examples. This resultant (hybrid) approach results from the first part of the 1st Law. It seeks to achieve effective and responsive coordination across several organizational contexts, especially in competitive environments.
   b) Governance capabilities [G] positively and partially mediate the relation between agile capabilities [A] and business operations [B] ($H_{16}$). Based on those results, we can also imply that the effect of agile capabilities [A] on business operations [B] has a better result when applied through the mediation of governance capabilities [G] (1st Law) than when applied directly upon [B] (2nd Law). It is closely related to the perception that "agility [A] can generate worrisome results without control [G]", assigning the required importance to the system responsible for the sense, response, and coordination of the entire organizational body: the steering system.

2) **eight hypotheses were supported in some scenarios** ($H_5$, $H_6$, $H_7$, $H_9$, $H_{10}$, $H_{11}$, $H_{12}$, and $H_{13}$), such as:
   a) The effects of moderator factors [M] have a negative influence on business operations [B] ($H_5$); and
   b) The agile capabilities [A] have a positive influence on the effects of moderator factors [M] ($H_{12}$).

3) **only three hypotheses were not supported in any scenario** ($H_8$, $H_{14}$, and $H_{15}$):
   a) The effects of moderator factors [M] have a negative influence on governance capabilities [G] ($H_8$);
   b) The governance capabilities [G] have a positive influence on the effects of moderator factors [M] ($H_{14}$); and,
   c) The governance capabilities [G] have a positive influence on the effects of environmental factors [E] ($H_{15}$).

Our results point to theory verisimilitude in support of hypotheses related to its key constructs: Agile capabilities [A], Governance capabilities [G], Business Operations [B], Value Delivery [R]; and core assumptions of the emerging theory in every theoretical scenario (see Section 5.1).

Returning to our original research question, "*Are the key hypotheses and core assumptions of the Agile Governance Theory supported by practitioners' experience?*" the empirical evidence has shown that the hypotheses related to key constructs and core assumptions of the emerging theory ($H_1$, $H_2$, $H_3$, $H_4$, and $H_{16}$) were **supported in all scenarios** in the inquiry, answering positively to the question. The hypotheses derived from the theory address AGT assessment. At the same time, our research question addresses the study's aim. We can claim that we have tested the various hypotheses that underpin the theory in practice and that the emerging theory's key constructs and core assumptions have been supported in every theoretical scenario tested.

### 6.2. Contributions and implications to theory

Our key contribution is assessing the *Agile Governance Theory,* advancing on what appears to be the first theory-building study about agile governance phenomena[25], combining the Agile and Lean philosophies with the Governance approach.

Because of the empirical study described in this report, we are attaining a *plausible* theory to describe and analyze the agile governance phenomena and the findings generated. The lessons learned by the present study can help design further studies for scrutinizing the AGT to reach a *trustworthy* theory (see Sections 5.1 and 5.4). Besides, the knowledge shared here can contribute to methodological tooling to help evaluate other theories.

The results from this research support the verisimilitude of previous studies allowing scholars and practitioners (i) to achieve a better characterization and differentiation between an agile governance approach and a specific agile approach (e.g., agile software development); as well as (ii) an evidence-based theory leading to improved plausibility, in aspects such as 1st Law, 2nd Law, 6th Law, and theory's premises published in (Luna et al., 2015). Based on the results of mediation analysis from $H_{16}$, (iii) we can also imply that the effect of Agility [A] on Business Operations [B] has a higher intensity when applied by the mediation of Governance capabilities [G] than when it is applied directly upon Business Operations [B].

### 6.3. Contributions and implications for practice

The findings of this study also support the plausibility of the conceptualization and characterization of agile governance as socio-technical phenomena positioned in a *chaordic* range between the innovation and emergent practices from agile and lean philosophies and the best practices employed as well as demanded by the governance issues. Our findings support the Agile Governance Manifesto foundations published in (Luna et al., 2016), considering the congruence of the six meta-principles, nine meta-values organized as a behavioral instrument to assist practitioners in applying agile governance in concrete organizational contexts.

---

[25] Considering we are studying intensely this area, at least for thirteen years, and we did not find any.

Regarding the practical implications of our findings, this study is a step toward testing AGT, meaning positive progress to help people and organizations improve teamwork and their capabilities of sense and response. Regardless of its size, area of expertise, or experience level, any organization may use AGT in its teams to achieve these benefits. However, AGT application is even more significant in competitive organizational environments. Likewise, when organizations need to grow and flourish, evolve to a new level, enter new markets, or face new challenges, the complexity of hierarchy and blurred decision channels can jeopardize their aims. In these cases, AGT might be helpful in establishing new strategies, governance structures, and control mechanisms without ignoring the need to be agile.

Finally, the assessment of the emergent theory to describe and analyze agile governance can be a precious contribution to establishing a unified view of the dynamic of agile governance practice, promoting an active discourse among scholars and practitioners, and driving them to act.

### 6.4. Future work

Our scientific study highlights where more research is required, and the following research topics could help to build a broader and more precise picture of the study phenomena:

(1) Further studies are needed to evaluate issues raised by the data gathered from participants' responses, such as (i) perception of "chaos and order scale" in their organizational contexts and its influence over agile governance practice; (ii) understanding of the participant's perception of some empirical indicators like Influence of regulatory institutions ($x_2$), Influence of competitiveness ($x_3$), Leadership inadequacy ($y_2$), and Compliance ($y_{13}$); to better explain the behavior of some theoretical constructs such as *Effects of environmental factors* [E], *Effects of moderator factors* [M], and *Governance capabilities* [G]; as well as (iii) further test the assumptions discussed in sections 4.2, 4.3.2, and 4.3.3. For instance: it would be a relevant topic for further investigation to *"understand how the 'nature of every project' influences the 'chaos and order perception' for each organizational context"*.

(2) We might replicate our study using case studies or action research to test and refine the theory. The results of these studies can be cross-checked and analyzed. New scenarios can be identified and tested to strengthen or refute behaviors predicted by the AGT.

(3) Connecting the theory with the real world, and identifying which frameworks, models, and applications help organizations employ agile governance. For instance, it seems relevant to evaluate how widely adopted frameworks for governance, such as SAFe (Razzak et al., 2018), are in connection with the AGT's analysis and description in each organizational context provided by the theory.

(4) It would be auspicious to perform an extension of the systematic literature review published in (Luna et al., 2014) to update the body of work for further studies on this topic.

**Acknowledgments**

We applied the SDC approach to the sequence of authors (Tscharntke, Hochberg, Rand, Resh, & Krauss, 2007). We are very grateful to Professor Philippe Kruchten from The University of British Columbia (UBC), for his priceless contributions and support throughout this research. Also, we are thankful to Sarah Beecham, a senior researcher from The Irish Software Research Centre (LERO), for her valuable contributions during the development and refinement of this report. The authors acknowledge CAPES, Brazil's Science without Borders Program, CNPq, and ATI-PE, the research support. Special thanks to Luciano José de Farias Silva and the FREVO[26] team for their valuable contributions. We were also grateful to Professor Júlio César Ferro de Guimarães from the Management Sciences Department at the Federal University of Pernambuco (UFPE), Professor Gauss Moutinho Cordeiro from the Statistics Department at the UFPE, Professor Renata Maria Cardoso Rodrigues de Souza from Computer Centre (CIn) at the same university, for their valuable feedback and considerations about the statistic approach carried out by this research. We are very thankful to all the authors, scholars, and practitioners with whom we have contacted and who participated in the several stages of this research, in which worthwhile contributions were instrumental in the outcome of this work. Also, we would like to thank: PGE-PE, The University of British Columbia (UBC) and St. John's College (SJC), the Department of Management Sciences (DCA-UFPE), Nucleus of Studies and Research in Information Systems (NEPSI), Project Research Group (GP2) at CIn-UFPE, Telehealth Nucleus (NUTES-UFPE), Clinics Hospital at UFPE (HC-UFPE), and the team of the Agile Governance Research Lab - AGRLab[27], where the unfolding of this work is already going on.

---

[26] FREVO — Fostering Research on managEment and InnoVatiOn is a concrete example of a self-organized multidisciplinary research team, and I am very proud and pleased to be part of this brotherhood. See more at: http://www.frevo.org.

[27] AGRLab channel for disseminating progress and research results: https://www.instagram.com/agilegovernancelab/

## APPENDIX A. THEORETICAL SCENARIOS EVOLVEMENT ($\varphi_1 .. \varphi_N$). ADAPTED FROM: (Luna et al., 2020).

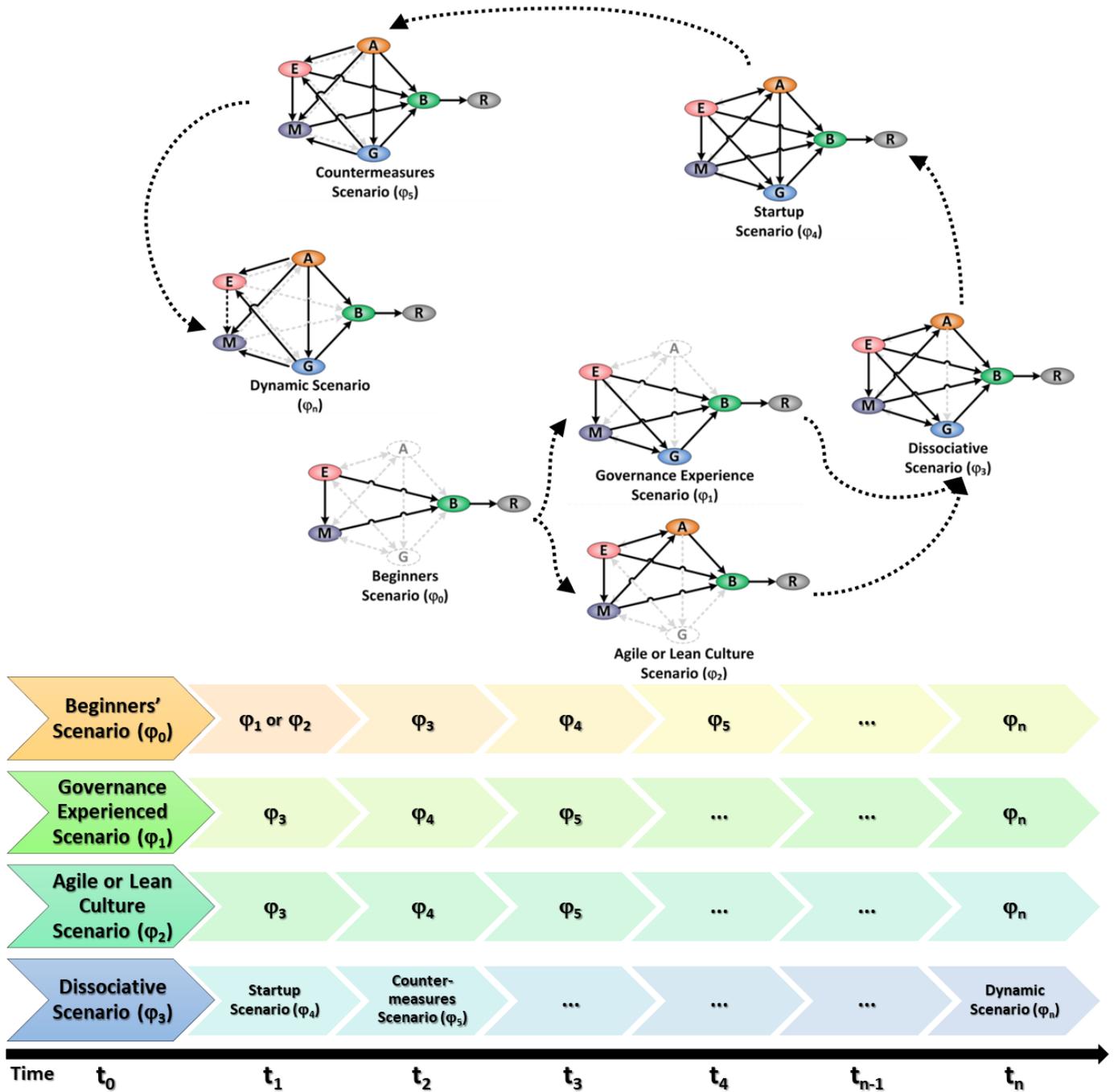

## APPENDIX B. Theory`s Empirical Indicators: summary of classification. ADAPTED FROM: (Luna et al., 2020).

*Empirical indicators*[28] identify operations that allow the researcher-theorist to measure the values of the constructs in the theoretical model. Besides the necessary accuracy for research, and due to the statistical methods planned to test the hypotheses (*Structural Equation Modeling* - SEM and *Confirmatory Factor Analysis* - CFA), based on Luna et al. (2020), we adopted more than one empirical indicator for each theory unit. The number of empirical indicators developed led us to a more consistent evidence-based study to assess the emerging theory.

The following table depicts the empirical indicators for each construct from the AGT adopted in this study. Those empirical indicators considered the feasibility of their measure for each distinct *theoretical scenario* where its respective construct was expected to manifest according to the theory. As a result, we have chosen the set of empirical indicators that they consider the best fit in most scenarios, totalizing 24 empirical indicators, as follows: (i) *five* empirical indicators for *Effects of environmental factors* [E] and *Moderator factors effects* [M]; (ii) *four* empirical indicators for *Agile capabilities* [A] and *Governance capabilities* [G]; and (iii) *three* empirical indicators for *Business Operations* [B] and *Value delivery* [R].

| #1[29] | #2[30] | Unit | Empirical Indicators | Metric description |
|---|---|---|---|---|
| $y_1$ | $x_1$ | [E] | Technological impact | [E] **as measured by** the degree of technological impact experienced by organizational context |
| $y_2$ | $x_2$ | | Influence of regulatory institutions | [E] **as measured by** the influence degree of regulatory institutions experienced by organizational context |
| $y_3$ | $x_3$ | | Influence of competitiveness | [E] **as measured by** the level of competitiveness experienced by organizational context |
| $y_4$ | $x_4$ | | Economy influence | [E] **as measured by** the level of economic influence experienced by organizational context |
| $y_5$ | $x_5$ | | Market turbulence | [E] **as measured by** the change rate in the environment experienced by organizational context |
| $y_6$ | $y_1$ | [M] | Organizational culture refractoriness | [M] **as measured by** the level of refractoriness of the organizational culture experienced by organizational context |
| $y_7$ | $y_2$ | | Leadership inadequacy | [M] **as measured by** the level of the inadequacy of leadership style experienced by organizational context |
| $y_8$ | $y_3$ | | Enterprise architecture inadequacy | [M] **as measured by** the level of the inadequacy of enterprise architecture experienced by organizational context |
| $y_9$ | $y_4$ | | Business model inadequacy | [M] **as measured by** the level of the inadequacy of business model experienced by organizational context |
| $y_{10}$ | $y_5$ | | Low-skilled people | [M] **as measured by** the level of lack of people qualification experienced by organizational context |
| $y_{11}$ | $y_6$ | [A] | Flexibility | [A] **as measured by** the level of flexibility experienced by organizational context |
| $y_{12}$ | $y_7$ | | Leanness | [A] **as measured by** the level of leanness experienced by organizational context |
| $y_{13}$ | $y_8$ | | Agility | [A] **as measured by** the level of agility experienced by organizational context |
| $y_{14}$ | $y_9$ | | Adaptability | [A] **as measured by** the level of adaptability experienced by organizational context |
| $y_{15}$ | $y_{10}$ | [G] | Strategic alignment | [G] **as measured by** the level of strategic alignment experienced by organizational context |
| $y_{16}$ | $y_{11}$ | | Decision making | [G] **as measured by** the level of decision-making experienced by organizational context |
| $y_{17}$ | $y_{12}$ | | Control | [G] **as measured by** the level of control experienced by organizational context |
| $y_{18}$ | $y_{13}$ | | Compliance | [G] **as measured by** the level of compliance experienced by organizational context |
| $y_{19}$ | $y_{14}$ | [B] | Business process-driven approach | [B] **as measured by** the degree of process-driven approach for permanent business aspects experienced by organizational context |
| $y_{20}$ | $y_{15}$ | | Projects driven approach | [B] **as measured by** the degree of project-based driven approach for transitory business aspects experienced by organizational context |
| $y_{21}$ | $y_{16}$ | | Best practices adoption | [B] **as measured by** the degree of best practices adoption for business, experienced by organizational context |
| $y_{22}$ | $y_{17}$ | [R] | Utility for product or service | [R] **as measured by** the grade of utility embedded in products or services experienced by organizational context |
| $y_{23}$ | $y_{18}$ | | Warranty for product or service | [R] **as measured by** the grade of warranty embedded in products or services, experienced by organizational context |
| $y_{24}$ | $y_{19}$ | | Time-to-market for product or service | [R] **as measured by** the degree of time-to-market of products and services experienced by organizational context |

---

[28] They are the actual instruments, experimental conditions and procedures that are used to observe or measure the concepts of middle-range theory (Dubin, 1978).

[29] For theoretical scenarios where **there are no** *independent manifest variables* ($\varphi_5$ and $\varphi_n$).

[30] For theoretical scenarios where **there are** *independent manifest variables* ($\varphi_0$, $\varphi_1$, $\varphi_2$, $\varphi_3$, and $\varphi_4$).

## APPENDIX C. Theory's Propositions: summary of classification.

| ID | Proposition | Proposition Statement | Traceability[31] |
|---|---|---|---|
| $P_1$* | Lifecycle | During the theory application, the [values of the] agile capabilities [A] and governance capabilities [G] will increase. | L1, L2, S2, S3, S4, S5 |
| $P_2$* | Business agility | If the agile capabilities [A] and governance capabilities [G] are high, then the business operations [B] [performance] will increase. | L1, L2, L5, S2 |
| $P_3$* | Value delivery | If the business operations [B] [performance] increase, then value delivery [R] will increase. | L6 |
| $P_4$* | Countermeasures | If the agile capabilities [A] and governance capabilities [G] are high, then the effects of environmental factors [E] and the effects of moderator factors [M] will decrease. | L3, L4, L5 |
| $P_5$* | Fewer effects | If the effects of environmental factors [E] and the effects of moderator factors [M] are low, then the business operations [B] [performance] will increase. | L3, L4, L5 |
| $P_6$ | Lethargy | Suppose the effects of environmental factors [E] and effects of moderator factors [M] are high. In that case, the business operations [B] [performance] will decrease, which can lead the whole system to a state of "*lethargy*" [S1]. | L3, L4, S1, S3 |
| $P_7$ | Sustainability & Competitiveness | The system state "*business agility*" [S2] will precede the system state [organizational] "*sustainability*" [S3] and "*competitiveness*" [S4]. | S2, S3, S4 |
| $P_8$ | Awareness or vitality | Suppose a balance between [organizational] "*sustainability*" [S3] and "*competitiveness*" [S4] is attained and maintained persistently (for a time enough to its institutional internalization). In that case, the organizational context goes into a state of [organizational] "*awareness*" [S5], achieving the whole system its maximum performance. | S3, S4, S5 |
| $P_9$ | Pre-theory states | All *pre-theory macro-system states* will precede the *theory of macro-system states*. | MS1, MS2, MS3, MS4, MS5, MS6, MS7 |
| $P_{10}$ | Internalizing | The macro-system state "*Startup*" [MS5] will precede the macro-system state "*Conscious Agile Governance*" [MS6]. | MS5, MS6 |
| $P_{11}$ | Quantum | The macro-system state "*Conscious Agile Governance*" [MS6] will precede the macro-system state "*Unconscious Agile Governance*" [MS7]. | MS6, MS7 |

## APPENDIX D. Theory's Hypotheses: summary of classification.

The *propositions* of a theory are a logical truth, a statement about the system in operation (derived from the logic underlying the theory) seeking to generate predictions, explanations, or descriptions about the observed phenomenon. At the same time, a *hypothesis* is a scientific instrument for disassembling a proposition to make it viable for testing in the real world (Dubin, 1978). The eleven *propositions* Luna et al. (2020) developed for AGT are depicted in APPENDIX C. The sixteen *hypotheses* designed by them to test the theory are characterized in the following table.

| ID | Category | Hypothesis statement | Traceability[32] |
|---|---|---|---|
| $H_1$ | Agile governance | Agile capabilities [A] positively influence governance capabilities [G]. | P1, P2, L1 |
| $H_2$ | Agile governance | The governance capabilities [G] positively influence business operations [B]. | P1, P2, L1 |
| $H_3$ | Specific agility | Agile capabilities [A] positively influence business operations [B]. | P1, P2, L2 |
| $H_4$ | Value delivery | The business operations [B], under the influence of agile capabilities [A] and governance capabilities [G], have a positive influence on value delivery [R]. | P1, P2, P3, L1, L6 |
| $H_5$ | Moderator factors effects | The effects of moderator factors [M] negatively influence business operations [B]. | P4, P5, P6, L3 |
| $H_6$ | Environmental factors effects | The effects of environmental factors [E] have a *negative influence* on business operations [B]. | P4, P5, P6, L4 |
| $H_7$ | Moderator factors effects | The effects of moderator factors [M] have a *negative influence* on agile capabilities [A]. | P4, P5, P6, L3 |
| $H_8$ | *Moderator factors effects* | *The effects of moderator factors [M] negatively influence governance capabilities [G].* | *P4, P5, P6, L3* |
| $H_9$ | Environmental Factors effects | The effects of environmental factors [E] have a *negative influence* on agile capabilities [A]. | P4, P5, P6, L4 |
| $H_{10}$ | Environmental factors effects | The effects of environmental factors [E] have a *negative influence* on governance capabilities [G]. | P4, P5, P6, L4 |
| $H_{11}$ | Environmental factors effects | The effects of environmental factors [E] have a positive influence on the effects of moderator factors [M]. | P4, P5, P6, L4 |
| $H_{12}$ | Sustainability | The agile capabilities [A] positively influence the effects of moderator factors [M]. | P2, P5, P7, L5 |
| $H_{13}$ | Competitiveness | The agile capabilities [A] positively influence environmental factors' effects [E]. | P2, P5, P7, L5 |
| $H_{14}$ | *Sustainability* | *The governance capabilities [G] positively influence the effects of moderator factors [M].* | *P2, P5, P7, L5* |
| $H_{15}$ | *Competitiveness* | *The governance capabilities [G] positively influence the effects of environmental factors [E].* | *P2, P5, P7, L5* |
| $H_{16}$ | Agile Governance mediation | Governance capabilities [G] positively and partially mediate the relation between agile capabilities [A] and business operations [B]. | P1, P2, L1 |

---

[31] Legend: Based on L – Law, MS – Macro-system state, or S – System state from Agile Governance Theory. For Instance, the P2 is based on the 1st, 2nd and 5th Laws of Agile Governance Theory.

[32] Legend: L – Law, P – Proposition.

Please cite as:

Luna, A. J. H. de O., & Marinho, M. L. M. (2023). *Multi-Scenario Empirical Assessment of Agile Governance Theory: A Technical Report*. Agile Governance Research Lab (AGRLab). DCA/CCSA-UFPE. Recife-PE. Brazil. Retrieved from https://www.agilegovernance.org/agt

**APPENDIX E.** AGT theoretical framework for testing: Models derived from the theoretical scenarios of AGT, considering: (A) Adjusted models, and (B) Models standardized estimates.

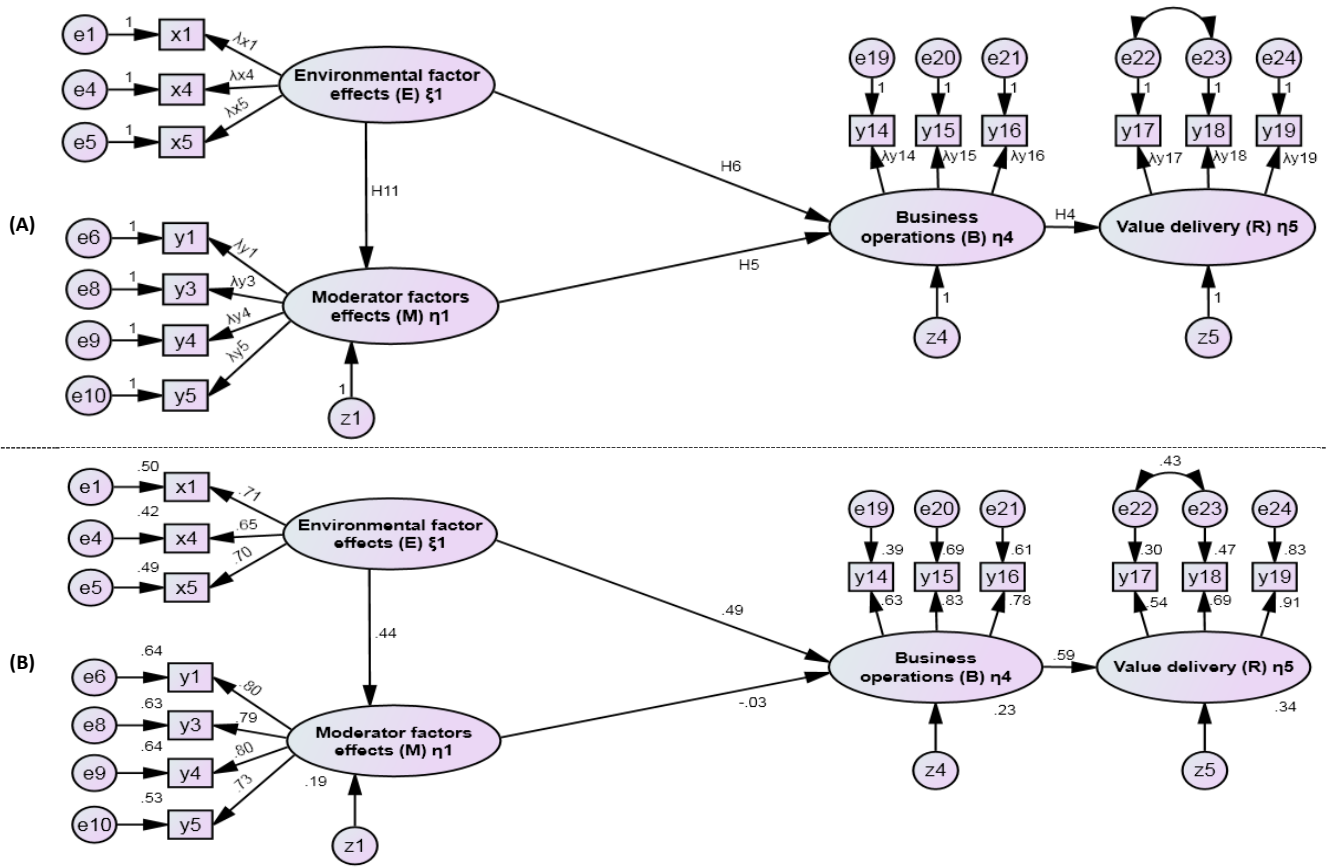

**Fig. E. 1.** SEM Theoretical model based on the *Beginners Scenario* ($\varphi_0$): (A) Adjusted model, (B) Estimation results.

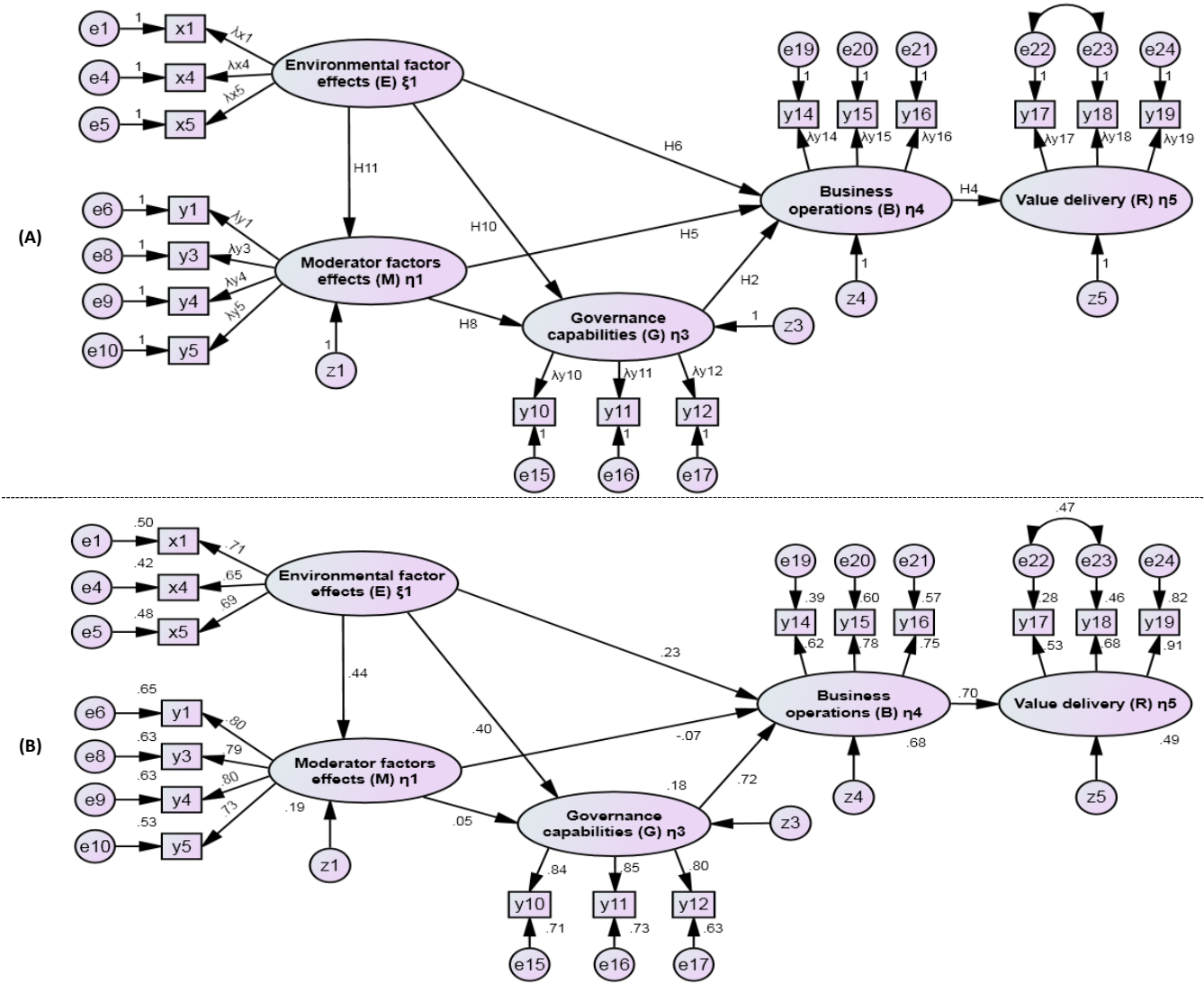

**Fig. E. 2. SEM Theoretical model based on the *Governance experience Scenario* ($\varphi_1$): (A) Adjusted model, (B) Estimation results.**

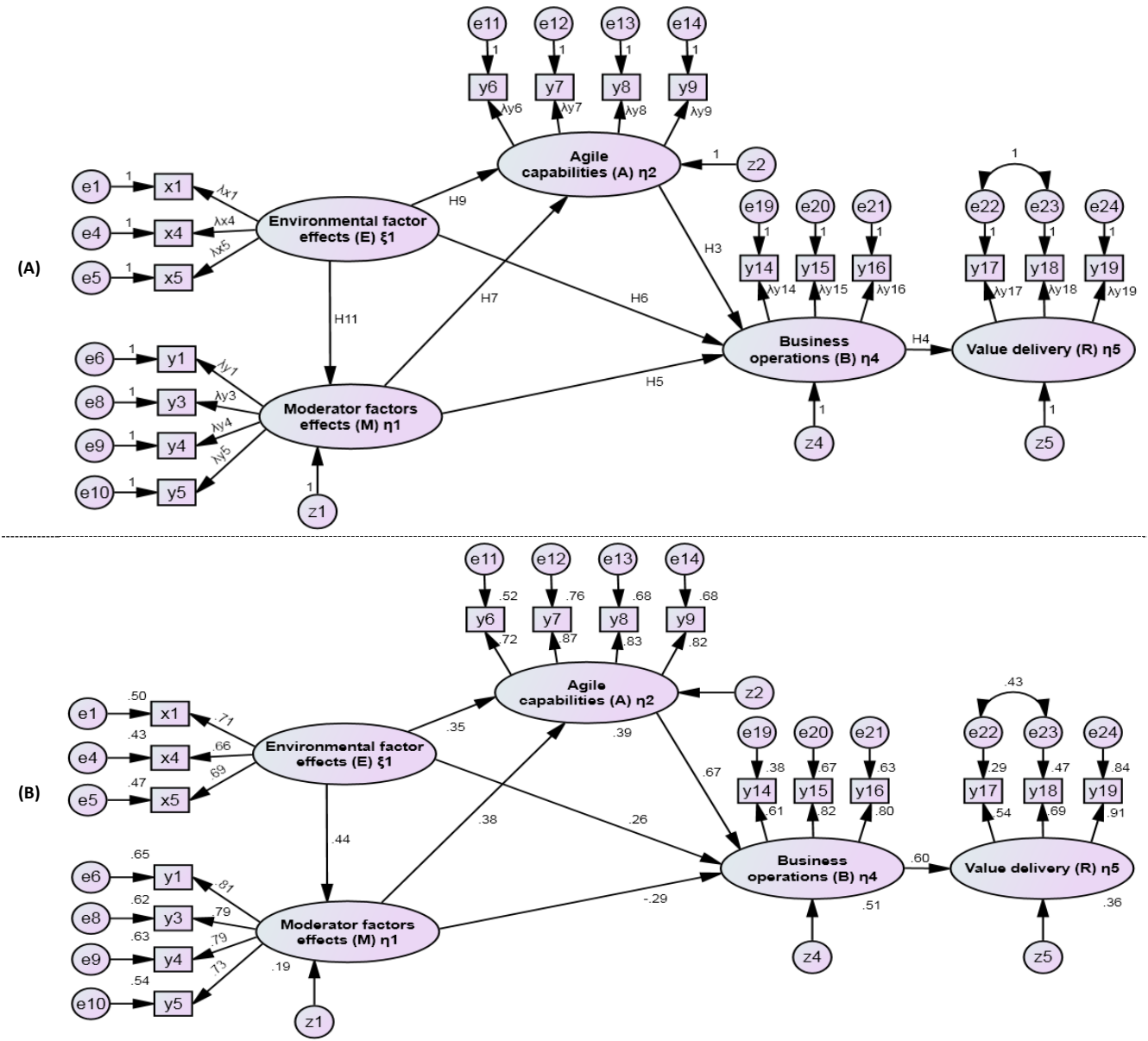

**Fig. E. 3.** SEM Theoretical model based on the *Agile experience Scenario* ($\varphi_2$): (A) Adjusted model, (B) Estimation results.

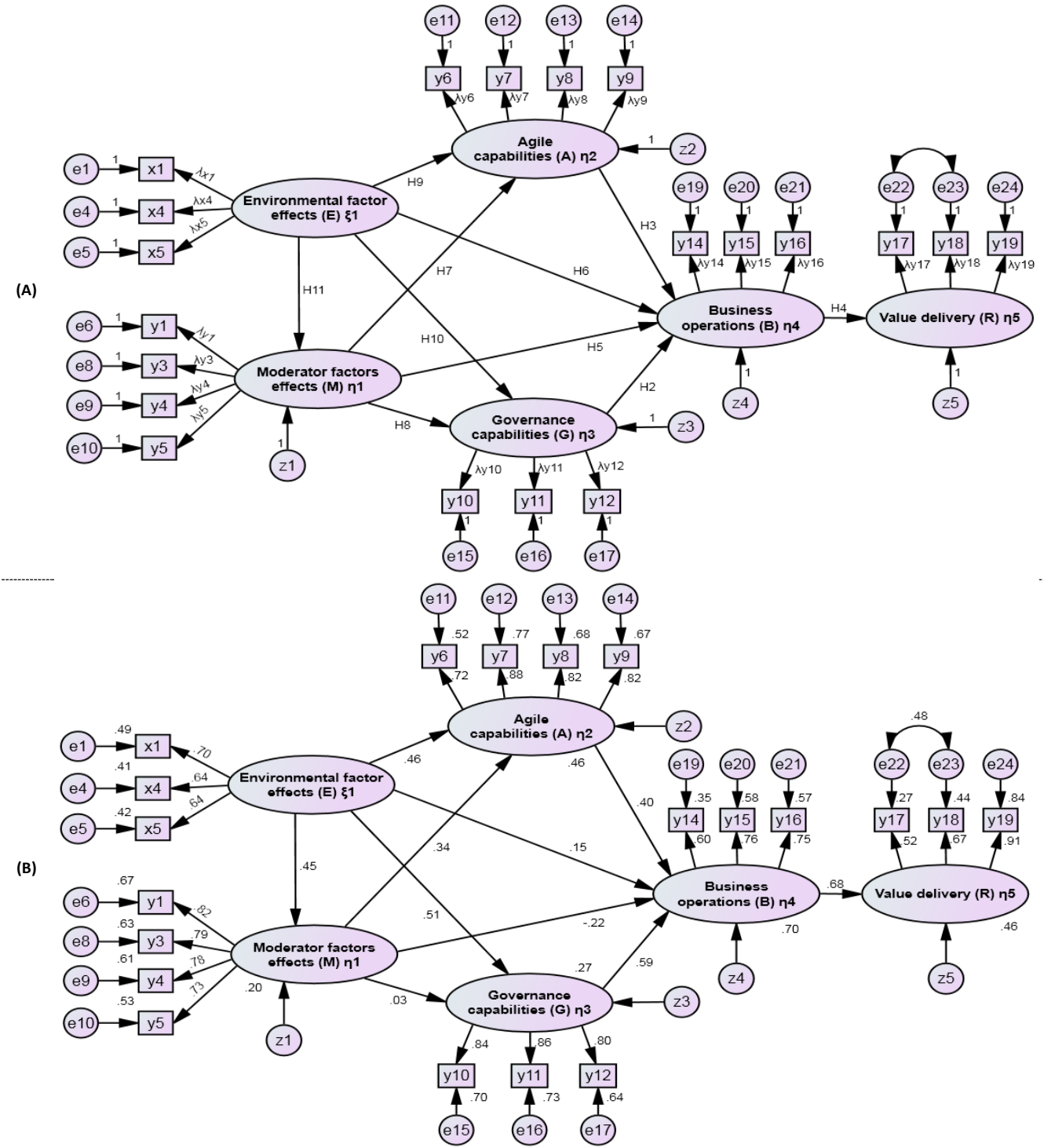

**Fig. E. 4.** SEM Theoretical model based on the *Dissociative Scenario* (φ₃): (A) Adjusted model, (B) Estimation results.

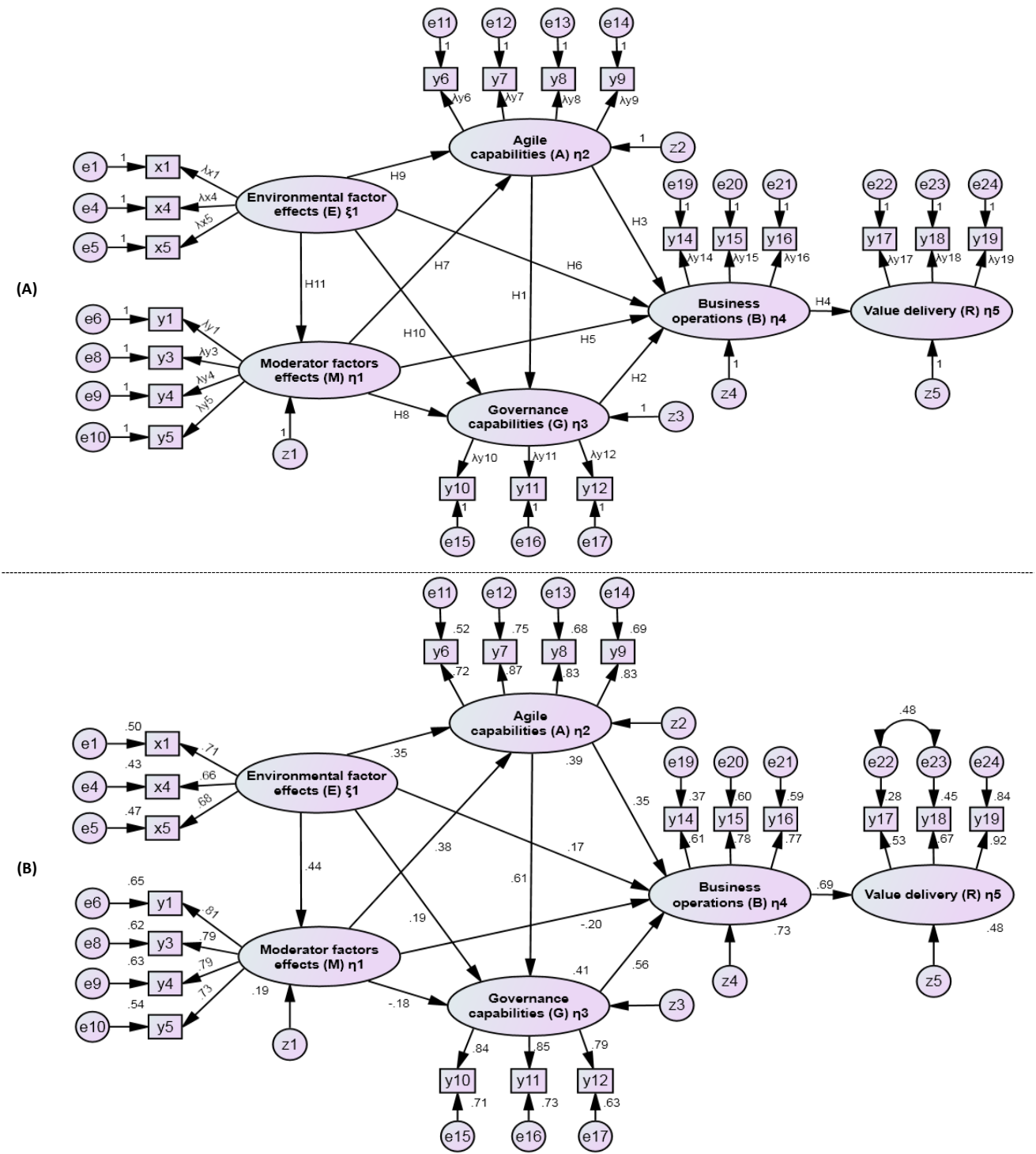

**Fig. E. 5.** SEM Theoretical model based on the *Startup Scenario* ($\varphi_4$): (A) Adjusted model, (B) Estimation results.

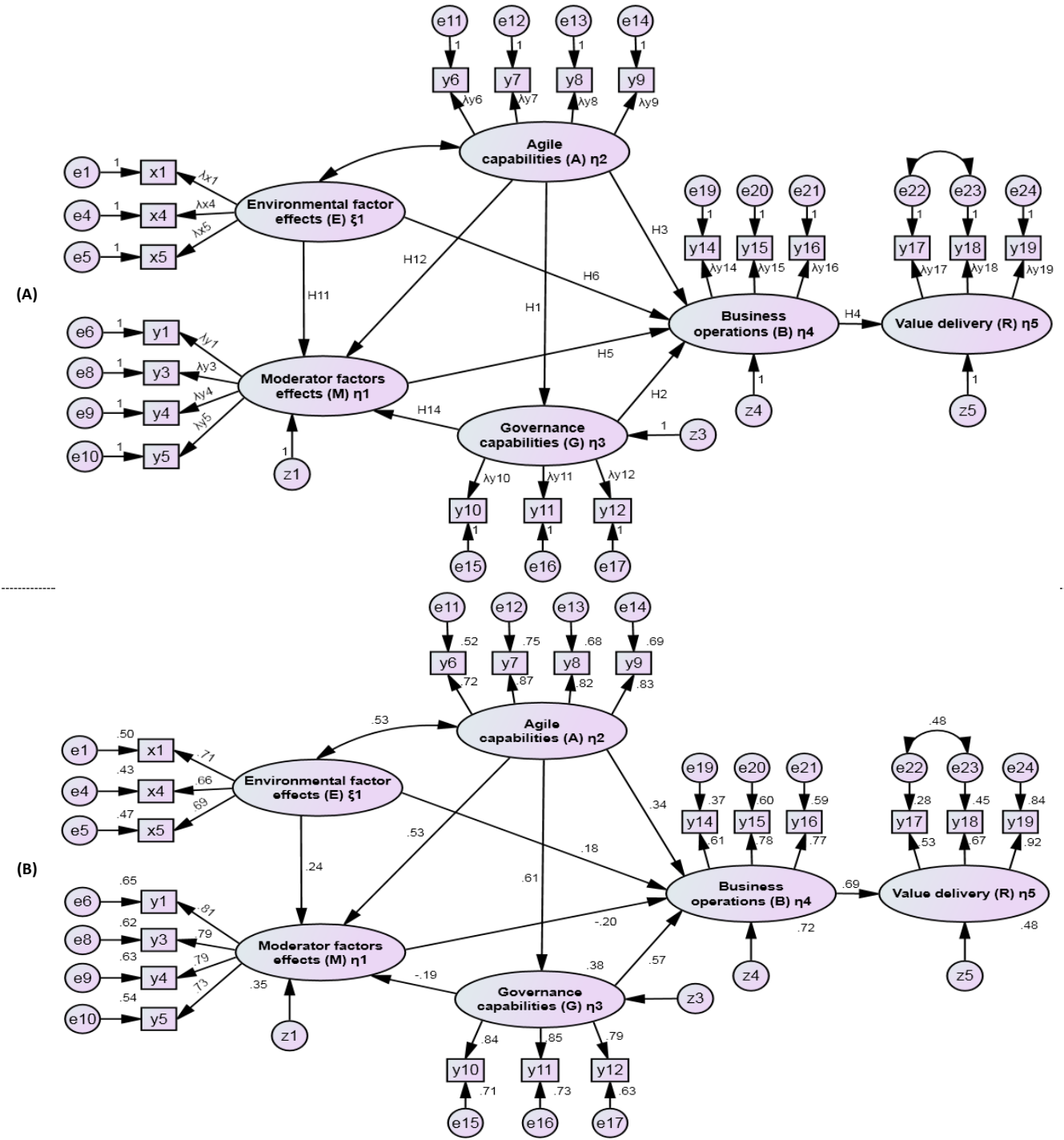

**Fig. E. 6.** SEM Theoretical model based on the *Sustainability Scenario* ($\varphi_{5'}$): (A) Adjusted model, (B) Estimation results.

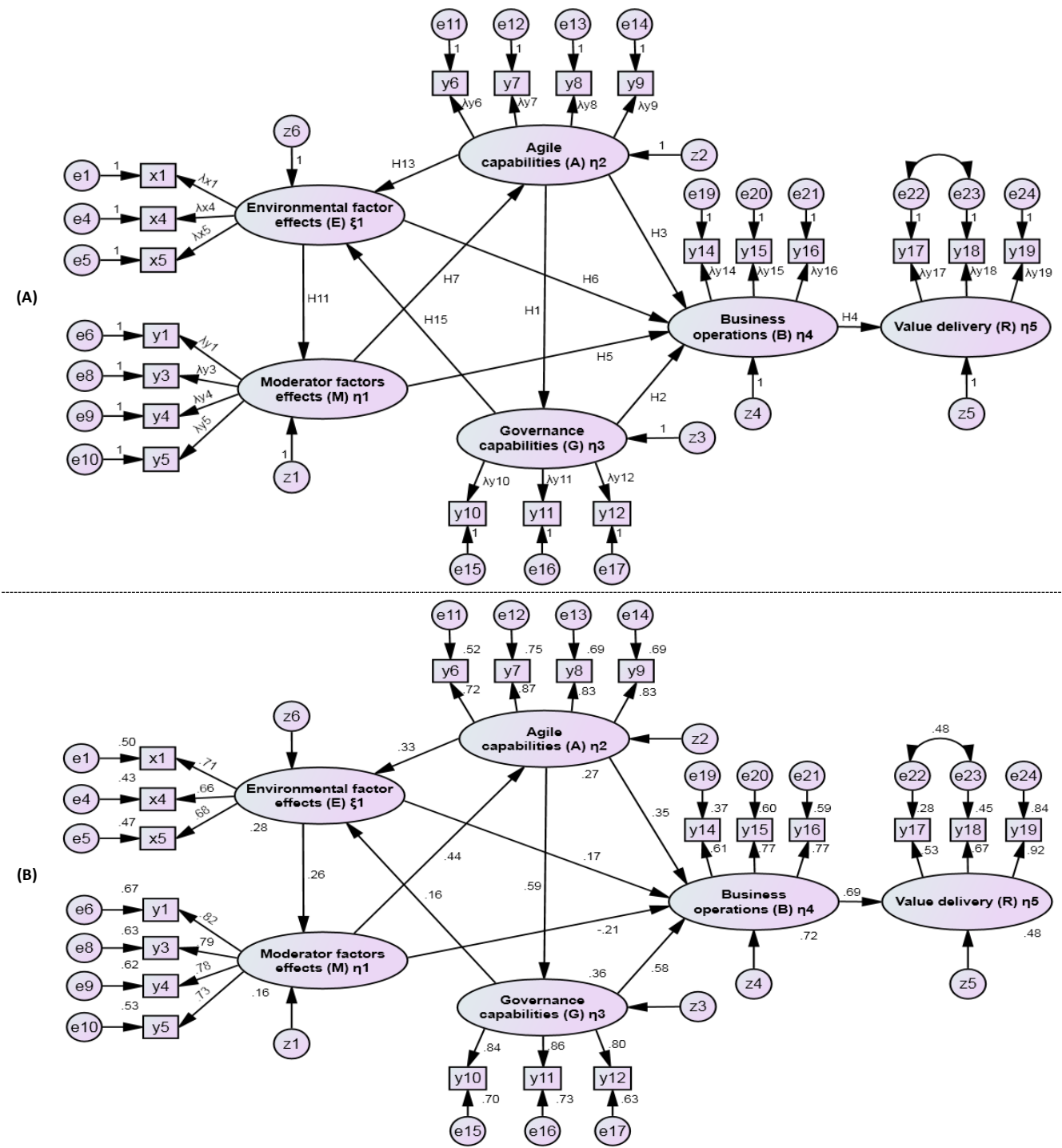

**Fig. E. 7.** SEM Theoretical model based on the *Competitiveness Scenario* ($\varphi_{5"}$): (A) Adjusted model, (B) Estimation results.

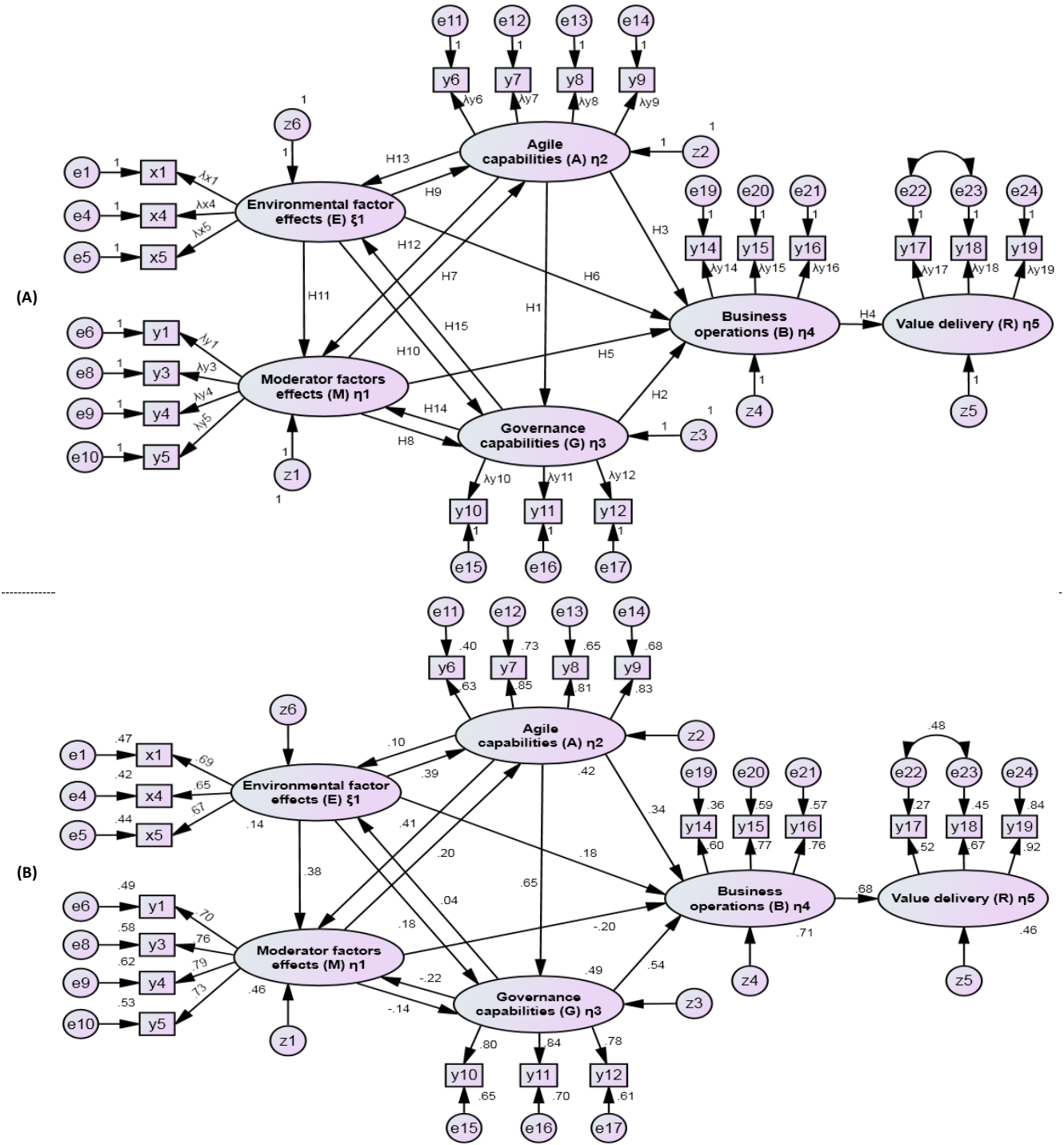

**Fig. E. 8.** SEM Theoretical model based on the *Startup Scenario* ($\varphi_4$): (A) Adjusted model, (B) Estimation results.

# APPENDIX F. AGT's Equations based on the Dynamic scenario ($\varphi_n$).

| (I) Structural sub-model: | |
|---|---|
| $E = H_{13}A + H_{15}G + z_1$ | **Equation F.1** – Effects of external environmental factors [E]. |
| $M = H_{11}E + H_{12}A + H_{14}G + z_2$ | **Equation F.2** – Effects of internal moderator factors [M]. |
| $A = H_9E + H_7M + z_3$ | **Equation F.3** – Agile capabilities [A]. |
| $G = H_{10}E + H_8M + H_1A + z_4$ | **Equation F.4** – Governance capabilities [G]. |
| $B = H_6E + H_5M + H_3A + H_2G + z_5$ | **Equation F.5** – Business operations [B]. |
| $R = H_4B + z_6$ | **Equation F.6** – Value delivery [R]. |
| $\therefore R = H_4(H_6E + H_5M + H_3A + H_2G + z_5) + z_6$ | **Equation F.7** – General Equation of the Structural theoretical model. |

| (1) Measurement sub-model from the construct Effects of external environmental factors [E]: | (2) Measurement sub-model from the construct Effects of internal moderator factors [M]: |
|---|---|
| $y_1 = \lambda_{y_1}E + e_1$ | $y_6 = \lambda_{y_6}M + e_6$ |
| $y_2 = \lambda_{y_2}E + e_2$ | $y_7 = \lambda_{y_7}M + e_7$ |
| $y_3 = \lambda_{y_3}E + e_3$ | $y_8 = \lambda_{y_8}M + e_8$ |
| $y_4 = \lambda_{y_4}E + e_4$ | $y_9 = \lambda_{y_9}M + e_9$ |
| $y_5 = \lambda_{y_5}E + e_5$ | $y_{10} = \lambda_{y_{10}}M + e_{10}$ |

| (3) Measurement sub-model from the construct Agile capabilities [A]: | (4) Measurement sub-model from the construct Governance capabilities [G]: |
|---|---|
| $y_{11} = \lambda_{y_{11}}A + e_{11}$ | $y_{15} = \lambda_{y_{15}}G + e_{15}$ |
| $y_{12} = \lambda_{y_{12}}A + e_{12}$ | $y_{16} = \lambda_{y_{16}}G + e_{16}$ |
| $y_{13} = \lambda_{y_{13}}A + e_{13}$ | $y_{17} = \lambda_{y_{17}}G + e_{17}$ |
| $y_{14} = \lambda_{y_{14}}A + e_{14}$ | $y_{18} = \lambda_{y_{18}}G + e_{18}$ |

| (5) Measurement sub-model from the construct Business operations [B]: | (6) Measurement sub-model from the construct Value delivery [R]: |
|---|---|
| $y_{19} = \lambda_{y_{19}}B + e_{19}$ | $y_{22} = \lambda_{y_{22}}R + e_{22}$ |
| $y_{20} = \lambda_{y_{20}}B + e_{20}$ | $y_{23} = \lambda_{y_{23}}R + e_{23}$ |
| $y_{21} = \lambda_{y_{21}}B + e_{21}$ | $y_{24} = \lambda_{y_{24}}R + e_{24}$ |

# APPENDIX G. Survey Questionnaire List.

| #Q | Group[33] | #1[34] | #2[35] | Unit | Empirical Indicator/ Variable | Question | Scale[36] |
|---|---|---|---|---|---|---|---|
| 1 | M | - | - | - | Language | *Which language do you prefer to answer this survey? | S |
| 2 | M | - | - | - | Organizational context | *Which of these organizational contexts best characterizes the background experience you will use to answer the questions of this survey? IMPORTANT: If you can contribute in more than one context, please answer the following question. | S |
| 3 | M | - | - | - | Other(s) Organizational context(s) | If you can respond to this questionnaire in more than one organizational context, please mark below which are they, and we will send another invitation to your email address. IMPORTANT: We must ensure that you keep in mind only one organizational context (at a time) when responding to this questionnaire. | S |
| 4 | M | - | - | - | Chaos & Order | *Based on the referential established in question 2 (Q2), how do you better characterize the chosen organizational context on a scale from chaos to order? | L |
| 5 | T | $y_1$ | $x_1$ | [E] | Technological impact | *You have experienced changes in business operations due to technological impact in the environment where the organizational context was inserted. | L |
| 6 | T | $y_2$ | $x_2$ | [E] | Influence of regulatory institutions | *You have experienced changes in business operations due to the influence of regulatory institutions from the environment where the organizational context was inserted. | L |
| 7 | T | $y_3$ | $x_3$ | [E] | Influence of competitiveness | *You have experienced changes in business operations due to the influence of competitors from the environment where the organizational context was inserted. | L |
| 8 | T | $y_4$ | $x_4$ | [E] | Economy influence | *You have experienced changes in business operations due to the economic influence upon the environment where the organizational context was inserted. | L |
| 9 | T | $y_5$ | $x_5$ | [E] | Market turbulence | * You have experienced changes in business operations due to the turbulence of the environment where the organizational context was inserted. | L |
| 10 | T | $y_6$ | $y_1$ | [M] | Organizational culture refractoriness | *You have experienced restraint or inhibition on business operations due to the influence of organizational culture. | L |
| 11 | T | $y_7$ | $y_2$ | [M] | Leadership inadequacy | *You have experienced restraint or inhibition of business operations due to the influence of leadership-related issues. | L |

---

[33] Legend for groups: M = participant's mindset, D = demography (sample qualification), T = theory assessment, A = study analysis.
[34] For theoretical scenarios where **there are no** independent manifest variables ($\varphi_5$ and $\varphi_n$).
[35] For theoretical scenarios where **there are** independent manifest variables ($\varphi_0$, $\varphi_1$, $\varphi_2$, $\varphi_3$, and $\varphi_4$).
[36] Legend for scales: S = single choice, M = multiple choice/select, L = 10-item Likert Scale, O = open-ended question.

| #Q | Group[33] | #1[34] | #2[35] | Unit | Empirical Indicator/Variable | Question | Scale[36] |
|---|---|---|---|---|---|---|---|
| 12 | T | $y_8$ | $y_3$ | | Enterprise architecture inadequacy | *You have experienced restraint or inhibition on the business operations due to the influence of inadequacy of the enterprise architecture. | L |
| 13 | T | $y_9$ | $y_4$ | | Business model inadequacy | *You have experienced restraint or inhibition on the business operations due to the influence of the inadequacy of how the business is structured to create, deliver, and capture value. | L |
| 14 | T | $y_{10}$ | $y_5$ | | Low-skilled people | *You have experienced restraint or inhibition of business operations due to the influence of a lack of people qualifications. | L |
| 15 | T | $y_{11}$ | $y_6$ | | Flexibility | *You have experienced circumstances where the adjustment capability of business operations to handle an unexpected situation was essential to ensure the <organizational context> performance. | L |
| 16 | T | $y_{12}$ | $y_7$ | [A] | Leanness | *You have experienced circumstances where the capability "to do more with less" on business operations was essential to ensure the <organizational context> performance. | L |
| 17 | T | $y_{13}$ | $y_8$ | | Agility | *You have experienced circumstances where the capability "to react to changes faster than the rate of these changes" on business operations was essential to ensure the <organizational context> performance. | L |
| 18 | T | $y_{14}$ | $y_9$ | | Adaptability | *You have experienced circumstances where the capability "to adapt evolutionarily" to business operations was essential to ensure the <organizational context> performance. | L |
| 19 | T | $y_{15}$ | $y_{10}$ | | Strategic alignment | *You have experienced circumstances where the capability "to ensure a continuous strategic alignment" on business operations was essential to guarantee the achievement of the <organizational context> objectives. | L |
| 20 | T | $y_{16}$ | $y_{11}$ | | Decision making | *You have experienced circumstances where the capability "to ensure effective decision making" on business operations was essential to guarantee the achievement of the <organizational context> objectives. | L |
| 21 | T | $y_{17}$ | $y_{12}$ | [G] | Control | *You have experienced circumstances where the capability "to ensure the strategy accomplishment" on business operations was essential to guarantee the achievement of the <organizational context> objectives. | L |
| 22 | T | $y_{18}$ | $y_{13}$ | | Compliance | *You have experienced circumstances where the capability "to ensure regulatory compliance status" of business operations was essential to guarantee the achievement of the <organizational context> objectives. | L |
| 23 | T | $y_{19}$ | $y_{14}$ | | Business process-driven approach | *You have experienced circumstances where the capability "to establish and implement business processes" to business operations was essential to guarantee the continuity of supply of products and/or services. | L |
| 24 | T | $y_{20}$ | $y_{15}$ | [B] | Projects driven approach | *You have experienced circumstances where the capability "to establish and implement a project-based approach" for business operations was essential to guarantee the continuity of supply of products and/or services. | L |
| 25 | T | $y_{21}$ | $y_{16}$ | | Best practices adoption | *You have experienced circumstances where the capability "to establish and implement best practices" on business operations was essential to guarantee the continuity of supply of products and/or services. | L |
| 26 | T | $y_{22}$ | $y_{17}$ | | Utility for product or service | *You have experienced circumstances where the "embedding of utility concept for products or services" led to increasing value delivery for the target audience (customers, citizens, and others). | L |
| 27 | T | $y_{23}$ | $y_{18}$ | [R] | Warranty for product or service | *You have experienced circumstances where the "embedding of warranty concept for products or services" led to increasing value delivery for the target audience (customers, citizens, and others). | L |
| 28 | T | $y_{24}$ | $y_{19}$ | | Time-to-market for product or service | *You have experienced circumstances where the "development of mechanisms to reduce time-to-market of products or services" led to increasing value delivery for the target audience. | L |
| 29 | D | - | - | - | Work experience | *How long is your work experience? | S |
| 30 | D | - | - | - | Job position | *What is your current job position? | M |
| 31 | D | - | - | - | Education | *What is your level of education (completed)? | S |
| 32 | D | - | - | - | Governance experience | *How long have you been directly or indirectly involved in initiatives to support governance? | S |
| 33 | D | - | - | - | Agile/Lean experience | *How long have you been directly or indirectly participating in initiatives using a lean or agile mindset (principles, values, practices, and others)? | S |
| 34 | D | - | - | - | Organization size | *How would you rate the size of the company where you work (or have worked recently)? | S |
| 35 | D | - | - | - | Economy sector | *In which industry sector your organization (predominantly) operates? | M |
| 36 | D | - | - | - | Operating scale | *What is the best classification for the operation scale of the organization where you work? | S |
| 37 | A | - | - | - | Actual adoption | *The organization where I work (or I have worked) adopts an agile approach to handling governance issues. | L |
| 38 | A | - | - | - | Future adoption | *The organization where I work (or I have worked) has an interest in adopting (or continue adopting) an agile approach to handle governance issues. | L |
| 39 | A | - | - | - | Contribution | *"The development of a theory for extending the understanding about agile governance phenomena is a necessary contribution to industry and academy". How much do you agree with this statement? | L |
| 40 | A | - | - | - | Feedback | If you are interested in receiving the results of this research, then fill in the following fields. | O |
| 41 | A | - | - | - | Suggestions | Please fill in the field below if you have any suggestions to improve this questionnaire or the ongoing research. | O |

*These questions require an answer.

# APPENDIX H. Factor validity from Constructs.

| Constructs | Manifest variables (Empirical indicators) | | Standardized factor weights ($\lambda$) | Level of significance for regression weight ($p$) |
|---|---|---|---|---|
| **Effects of environmental factors [E]** | $x_1$ | Technological impact | 0.710 | N/A |
| | $x_4$ | Economic influence | 0.657 | *** |
| | $x_5$ | Market turbulence | 0.682 | *** |
| **Effects of moderator factors [M]** | $y_1$ | Organizational culture refractoriness | 0.808 | N/A |
| | $y_3$ | Enterprise architecture inadequacy | 0.788 | *** |
| | $y_4$ | Business model inadequacy | 0.794 | *** |
| | $y_5$ | Low-skilled people | 0.734 | *** |
| **Agile capabilities [A]** | $y_6$ | Flexibility | 0.721 | N/A |
| | $y_7$ | Leanness | 0.866 | *** |
| | $y_8$ | Agility | 0.826 | *** |
| | $y_9$ | Adaptability | 0.831 | *** |
| **Governance capabilities [G]** | $y_{10}$ | Strategic alignment | 0.844 | N/A |
| | $y_{11}$ | Decision making | 0.852 | *** |
| | $y_{12}$ | Control | 0.794 | *** |
| **Business operations [B]** | $y_{14}$ | Business process-driven approach | 0.612 | N/A |
| | $y_{15}$ | Projects driven approach | 0.777 | *** |
| | $y_{16}$ | Best practices adoption | 0.767 | *** |
| **Value delivery [R]** | $y_{17}$ | Utility for product or service | 0.528 | N/A |
| | $y_{18}$ | Warranty for product or service | 0.673 | *** |
| | $y_{19}$ | Time-to-market for product or service | 0.917 | *** |

\***$p$ <0.001**: this means that in the relation [$Construct \leftarrow Manifest\ variable$] the regression weight for $Manifest\ variable$ in the prediction of the $Construct$ is significantly different from zero at the 0.001 level (two-tailed).

**N/A:** The standard errors for some items that were not calculated were because their charges were set at 1.0 for model identification purposes and calculation.

## APPENDIX I. Hypothesis test for Startup Scenario ($\varphi_4$), (n = 118).

| | Hypotheses | Startup Scenario ($\varphi_4$) | Test | Comments |
|---|---|---|---|---|
| $H_1$ | Agile governance [G ← A] | Significant ($\beta = 0.606$, ***) | Supported | N/A |
| $H_2$ | Agile governance [B ← G] | Significant ($\beta = 0.557$, ***) | Supported | N/A |
| $H_3$ | Specific agility [B ← A] | Significant ($\beta = 0.349$, $p = 0.011$ *) | Supported | N/A |
| $H_4$ | Value delivery [R ← B] | Significant ($\beta = 0.691$, ***) | Supported | N/A |
| $H_5$ | Moderator factors effects [B ← M] | Non-significant ($\beta = -0.196$, $p = 0.066$) | Not supported | Even though we can identify the negative influence of [M] upon [B] in the current scenario, the hypothesis has not enough statistical significance. |
| $H_6$ | Environmental factors effects [B ← E] | Non-significant ($\beta = 0.169$, $p = 0.144$) | Not supported | Despite the hypothesis not having statistical significance, the influence identified from [E] upon [B] by the study is slight and positive. When we realize that not only threats (negative influence) from Environmental factors effects [E] can jeopardize Business operations [B], but also opportunities (positive influence) can potentiate it. The emerging theory supports this evidence. |
| $H_7$ | Moderator factors effects [A ← M] | Significant ($\beta = 0.384$, ***) | Supported | Despite the hypothesis having statistical significance, the influence identified from [M] upon [A] by the study is positive instead of negative, as depicted in the hypothesis statement. This result leads us to think whether: (1) it would be this one occurrence of the behavior changing from Moderator factors effect [M] into enabler ones, such as when it is empowered by the combination of agile and governance capabilities; or (2) whether, this result can be a consequence of the multiple organizational contexts shuffled into the data sample; or even, (3) this evidence can be an effect from the wording (quite generic questions) we have asked on the questionnaire. Regardless of these hypotheses, this denouement has to be further investigated in future studies. |
| $H_8$ | Moderator factors effects [G ← M] | Non-significant ($\beta = -0.184$, $p = 0.131$) | Not supported | Although we have identified the negative influence from [M] upon [G] in the current scenario, the hypothesis has not enough statistical significance. |
| $H_9$ | Environmental Factors effects [A ← E] | Significant ($\beta = 0.350$, $p = 0.005$ **) | Supported | Despite the hypothesis having statistical significance, the influence identified from [E] upon [A] by the study is positive instead of negative, as depicted in the hypothesis statement. When we realize that not only threats (negative influence) from Environmental factors effects [E] can disturb Agile capabilities [G] development, but also opportunities (positive influence) can leverage it. The emerging theory supports this evidence. |
| $H_{10}$ | Environmental factors effects [G ← E] | Non-significant ($\beta = 0.192$, $p = 0.131$) | Not supported | Despite the hypothesis not having statistical significance, the influence identified from [E] upon [G] by the study is slight and positive. When we realize that not only threats (negative influence) from Environmental factors effects [E] can disturb Governance capabilities [G] development, but also opportunities (positive influence) can leverage it. The emerging theory supports this evidence. |
| $H_{11}$ | Environmental factors effects [M ← E] | Significant ($\beta = 0.439$, ***) | Supported | N/A |
| $H_{12}$ | Sustainability [M ← A] | N/A | N/A | N/A |
| $H_{13}$ | Competitiveness [E ← A] | N/A | N/A | N/A |
| $H_{14}$ | Sustainability [M ← G] | N/A | N/A | N/A |
| $H_{15}$ | Competitiveness [E ← G] | N/A | N/A | N/A |

Where: "$\beta$" are "standardized regression weights", pointing out the factor validity between related variables [Y ← X], and "$p$" are "standard errors", pointing out the statistical significance from each relation ($\beta$).

* This means that in the relation [Y ← X], the regression weight for X in the prediction of Y is significantly different from zero at the **0.05** level (two-tailed).

** This means that in the relation [Y ← X], the regression weight for X in the prediction of Y is significantly different from zero at the **0.01** level (two-tailed).

*** ***p <0.001**, this means that in the relation [Y ← X], the regression weight for X in the prediction of Y is significantly different from zero at the 0.001 level (two-tailed).

N/A: When the hypothesis is not contemplated in the scenario.

# APPENDIX J. Quality adjustment indices from Agile Governance theoretical scenarios (n = 118).

| Group Analysis | Indices | Beginners' Scenario ($\varphi_0$) | Governance experienced Scenario ($\varphi_1$) | Agile or Lean culture Scenario ($\varphi_2$) | Dissociative Scenario ($\varphi_3$) | Startup Scenario ($\varphi_4$) | Sustainability Scenario ($\varphi_{5'}$) | Competitiveness Scenario ($\varphi_{5''}$) | Dynamic Scenario ($\varphi_n$) | Reference values for Analysis |
|---|---|---|---|---|---|---|---|---|---|---|
| *Fit tests* | Chi-square ($\chi^2$) | 58.980 | 142.640 | 112.520 | 223.829 | 202.300 | 203.838 | 204.705 | 218.480 | lower is better |
| | Degrees of freedom ($df$) | 61 | 96 | 112 | 159 | 158 | 159 | 159 | 158 | ≥1 |
| | p-value | 0.549 | 0.001 | 0.468 | 0.001 | 0.010 | 0.009 | 0.008 | 0.001 | >0.05 |
| *Absolute Indices* | Standardized Chi-square ($\chi^2/df$) | 0.967 | 1.486 | 1.005 | 1.408 | 1.280 | 1.282 | 1.287 | 1.383 | <3 |
| | Root Mean Square Error of Approximation (RMSEA) | 0.000 | 0.064 | 0.006 | 0.059 | 0.049 | 0.049 | 0.050 | 0.057 | <0.10 |
| | Goodness of Fit Index (GFI) | 0.928 | 0.874 | 0.901 | 0.844 | 0.857 | 0.856 | 0.856 | 0.848 | >0.90 |
| *Relative Indices* | Comparative Fit Index (CFI) | 1.000 | 0.945 | 0.999 | 0.945 | 0.962 | 0.962 | 0.961 | 0.948 | >0.90 |
| | Normed Fit Index (NFI) | 0.907 | 0.852 | 0.890 | 0.835 | 0.851 | 0.850 | 0.850 | 0.839 | >0.90 |
| | Tucker-Lewis Index (TLI) | 1.005 | 0.931 | 0.999 | 0.934 | 0.954 | 0.954 | 0.953 | 0.938 | >0.90 |
| *Parsimony Indices* | Parsimony GFI (PGFI) | 0.622 | 0.617 | 0.659 | 0.639 | 0.644 | 0.648 | 0.648 | 0.638 | >0.60 |
| | Parsimony CFI (PCFI) | 0.782 | 0.756 | 0.823 | 0.790 | 0.800 | 0.805 | 0.804 | 0.789 | >0.60 |
| | Parsimony NFI (PNFI) | 0.710 | 0.681 | 0.733 | 0.699 | 0.708 | 0.711 | 0.711 | 0.698 | >0.60 |
| | *Overall Goodness of Fit* | Good Fit | Acceptable fit | Acceptable fit | Acceptable fit | Acceptable fit | Acceptable fit | Acceptable fit | Acceptable fit | - |

# APPENDIX K. Hypothesis test for Agile Governance theoretical scenarios, (n = 118).

| | Hypotheses | Beginner Scenario ($\varphi_0$) | Governance experience Scenario ($\varphi_1$) | Agile or Lean culture Scenario ($\varphi_2$) | Dissociative Scenario ($\varphi_3$) | Startup Scenario ($\varphi_4$) | Sustainability Scenario ($\varphi_{5'}$) | Competitiveness Scenario ($\varphi_{5''}$) | Dynamic Scenario ($\varphi_n$) | Statistics | Analysis |
|---|---|---|---|---|---|---|---|---|---|---|---|
| $H_1$ | Agile governance [G ← A] | N/A | N/A | N/A | N/A | Significant ($\beta = 0.606$, ***) | Significant ($\beta = 0.613$, ***) | Significant ($\beta = 0.587$, ***) | Significant ($\beta = 0.652$, ***) | **Passed**:4/4 **Refuted**:0/4 | 100% success |
| $H_2$ | Agile governance [B ← G] | N/A | Significant ($\beta = 0.720$, ***) | N/A | Significant ($\beta = 0.589$, ***) | Significant ($\beta = 0.557$, ***) | Significant ($\beta = 0.568$, ***) | Significant ($\beta = 0.576$, ***) | Significant ($\beta = 0.540$, ***) | **Passed**:6/6 **Refuted**:0/6 | 100% success |
| $H_3$ | Specific agility [B ← A] | N/A | N/A | Significant ($\beta = 0.675$, ***) | Significant ($\beta = 0.399$, $p = 0.004$ **) | Significant ($\beta = 0.349$, $p = 0.011$ *) | Significant ($\beta = 0.338$, $p = 0.016$ *) | Significant ($\beta = 0.350$, $p = 0.009$ **) | Significant ($\beta = 0.335$, $p = 0.026$ *) | **Passed**:6/6 **Refuted**:0/6 | 100% success |
| $H_4$ | Value delivery [R ← B] | Significant ($\beta = 0.585$, ***) | Significant ($\beta = 0.702$, ***) | Significant ($\beta = 0.603$, ***) | Significant ($\beta = 0.677$, ***) | Significant ($\beta = 0.691$, ***) | Significant ($\beta = 0.692$, ***) | Significant ($\beta = 0.690$, ***) | Significant ($\beta = 0.682$, ***) | **Passed**:8/8 **Refuted**:0/8 | 100% success |
| $H_5$ | Moderator factors effects [B ← M] | Non-significant ($\beta = -0.027$, $p = 0.823$) | Non-significant ($\beta = -0.071$, $p = 0.461$) | Significant ($\beta = -0.288$, $p = 0.018$ **) | Significant ($\beta = -0.222$, $p = 0.041$ *) | Non-significant ($\beta = -0.196$, $p = 0.066$) | Non-significant ($\beta = -0.198$, $p = 0.066$) | Significant ($\beta = -0.212$, $p = 0.042$) | Non-significant ($\beta = -0.195$, $p = 0.100$) | **Passed**:3/8 **Refuted**:5/8 | 38% success |
| $H_6$ | Environmental factors effects [B ← E] | Significant ($\beta = 0.494$, $p = 0.002$ **) | Non-significant ($\beta = 0.232$, $p = 0.058$) | Non-significant ($\beta = 0.261$, $p = 0.052$) | Non-significant ($\beta = 0.150$, $p = 0.275$) | Non-significant ($\beta = 0.169$, $p = 0.144$) | Non-significant ($\beta = 0.181$, $p = 0.119$) | Non-significant ($\beta = 0.181$, $p = 0.138$) | Non-significant ($\beta = 0.178$, $p = 0.168$) | **Passed**:1/8 **Refuted**:7/8 | 13% success |
| $H_7$ | Moderator factors effects [A ← M] | N/A | N/A | Significant ($\beta = 0.383$, ***) | Significant ($\beta = 0.336$, $p = 0.003$ **) | Significant ($\beta = 0.384$, ***) | N/A | Significant ($\beta = 0.443$, ***) | Non-significant ($\beta = 0.202$, $p = 0.233$) | **Passed**:4/5 **Refuted**:1/5 | 80% success |

| | Hypotheses | Beginner Scenario ($\varphi_0$) | Governance experience Scenario ($\varphi_1$) | Agile or Lean culture Scenario ($\varphi_2$) | Dissociative Scenario ($\varphi_3$) | Startup Scenario ($\varphi_4$) | Sustainability Scenario ($\varphi_{5'}$) | Competitiveness Scenario ($\varphi_{5''}$) | Dynamic Scenario ($\varphi_n$) | Statistics | Analysis |
|---|---|---|---|---|---|---|---|---|---|---|---|
| $H_8$ | Moderator factors effects $[G \leftarrow M]$ | N/A | Non-significant ($\beta = 0.050$, $p = 0.682$) | N/A | Non-significant ($\beta = 0.030$, $p = 0.814$) | Non-significant ($\beta = -0.184$, $p = 0.131$) | N/A | N/A | Non-significant ($\beta = -0.144$, $p = 0.383$) | **Passed**:0/4 **Refuted**:4/4 | 0% success |
| $H_9$ | Environmental Factors effects $[A \leftarrow E]$ | N/A | N/A | Significant ($\beta = 0.351$, ***) | Significant ($\beta = 0.457$, $p = 0.001$ ***) | Significant ($\beta = 0.350$, $p = 0.005$ **) | N/A | N/A | Non-significant ($\beta = 0.385$, $p = 0.081$) | **Passed**:3/4 **Refuted**:1/4 | 75% success |
| $H_{10}$ | Environmental factors effects $[G \leftarrow E]$ | N/A | Significant ($\beta = 0.403$, $p = 0.003$ **) | N/A | Significant ($\beta = 0.509$, $p = 0.001$ ***) | Non-significant ($\beta = 0.192$, $p = 0.131$) | N/A | N/A | Non-significant ($\beta = 0.176$, $p = 0.436$) | **Passed**:2/4 **Refuted**:2/4 | 50% success |
| $H_{11}$ | Environmental factors effects $[M \leftarrow E]$ | Significant ($\beta = 0.438$, ***) | Significant ($\beta = 0.438$, ***) | Significant ($\beta = 0.439$, ***) | Significant ($\beta = 0.447$, ***) | Significant ($\beta = 0.439$, ***) | Non-significant ($\beta = 0.235$, $p = 0.070$) | Non-significant ($\beta = 0.263$, $p = 0.066$) | Significant ($\beta = 0.380$, $p = 0.027$ *) | **Passed**:6/8 **Refuted**:2/8 | 75% success |
| $H_{12}$ | Sustainability $[M \leftarrow A]$ | N/A | N/A | N/A | N/A | N/A | Significant ($\beta = 0.531$, ***) | N/A | Non-significant ($\beta = 0.411$, $p = 0.055$) | **Passed**:1/2 **Refuted**:1/2 | 50% success |
| $H_{13}$ | Competitiveness $[E \leftarrow A]$ | N/A | N/A | N/A | N/A | N/A | N/A | Significant ($\beta = 0.332$, $p = 0.033$ *) | Non-significant ($\beta = 0.103$, $p = 0.729$) | **Passed**:1/2 **Refuted**:1/2 | 50% success |
| $H_{14}$ | Sustainability $[M \leftarrow G]$ | N/A | N/A | N/A | N/A | N/A | Non-significant ($\beta = -0.191$, $p = 0.136$) | N/A | Non-significant ($\beta = -0.221$ $p = 0.246$) | **Passed**:0/2 **Refuted**:2/2 | 0% success |
| $H_{15}$ | Competitiveness $[E \leftarrow G]$ | N/A | N/A | N/A | N/A | N/A | N/A | Non-significant ($\beta = 0.164$, $p = 0.252$) | Non-significant ($\beta = 0.039$, $p = 0.896$) | **Passed**:0/2 **Refuted**:2/2 | 0% success |
| $H_{16}$ | Agile Governance mediation $\begin{bmatrix} A \\ \downarrow \searrow \\ G \rightarrow B \end{bmatrix}$ | N/A | N/A | N/A | N/A | • Sobel test: For $\alpha = 0.05$, $\|Z\| = 2.965 > Z_{0.975} = 1.96$, $p = 0.003$ **: Significant • Bootstrap test: For $p_{A \rightarrow B} = 0.001$ ***, $p_{G \rightarrow B} = 0.002$ ***: Significant | • Sobel test: For $\alpha = 0.05$, $\|Z\| = 3.339 > Z_{0.975} = 1.96$, $p = 0.001$ ***: Significant • Bootstrap test: For $p_{A \rightarrow B} = 0.001$ ***, $p_{G \rightarrow B} = 0.001$ ***: Significant | • Sobel test: For $\alpha = 0.05$, $\|Z\| = 3.325 > Z_{0.975} = 1.96$, $p = 0.001$ ***: Significant • Bootstrap test: For $p_{A \rightarrow B} = 0.001$ ***, $p_{G \rightarrow B} = 0.001$ ***: Significant | • Sobel test: For $\alpha = 0.05$, $\|Z\| = 2.829 > Z_{0.975} = 1.96$, $p = 0.005$ **: Significant • Bootstrap test: For $p_{A \rightarrow B} = 0.012$ **, $p_{G \rightarrow B} = 0.004$ ***: Significant | **Passed**:4/4 **Refuted**:0/4 | 100% success |
| | *Statistics* | **Passed**: 3/4 **Refuted**:1/4 | **Passed**: 4/7 **Refuted**:3/7 | **Passed**: 6/7 **Refuted**: 1/7 | **Passed**: 8/10 **Refuted**: 2/10 | **Passed**: 8/12 **Refuted**: 4/12 | **Passed**: 6/10 **Refuted**: 4/10 | **Passed**: 8/11 **Refuted**: 3/11 | **Passed**: 6/16 **Refuted**: 10/16 | - | - |
| | *Analysis* | 75% success | 57% success | 86% success | 80% success | 67% success | 60% success | 73% success | 38% success | - | - |
| | *Overall Goodness of Fit* | Good Fit | Acceptable fit | Acceptable fit | Acceptable fit | Acceptable fit | Acceptable fit | Acceptable fit | Acceptable fit | - | - |

Where: "$\beta$" are "standardized regression weights", pointing out the factor validity between related variables $[Y \leftarrow X]$, and "$p$" are "standard errors", pointing out the statistical significance from each relation ($\beta$).
* This means that in the relation $[Y \leftarrow X]$, the regression weight for $X$ in the prediction of $Y$ is significantly different from zero at the **0.05** level (two-tailed). ** This means that in the relation $[Y \leftarrow X]$, the regression weight for $X$ in the prediction of $Y$ is significantly different from zero at the **0.01** level (two-tailed). *** $p$ <**0.001**, this means that in the relation $[Y \leftarrow X]$, the regression weight for $X$ in the prediction of $Y$ is significantly different from zero at the 0.001 level (two-tailed).
N/A: When the hypothesis is not contemplated in the scenario.